\documentclass[table,twocolumn,mathline]{aastex631}
\usepackage[utf8]{inputenc}
\usepackage{rotating}
\usepackage{color}
\usepackage{graphicx}
\usepackage{subfigure}
\usepackage{array}
\usepackage{comment}
\usepackage{soulutf8} 
\usepackage{xcolor}
\usepackage{ulem}  

\DeclareUnicodeCharacter{02BC}{'}

\newcommand{\temponest}{\textsc{temponest}~}

\begin{document}
\shorttitle{M2 pulsars}
\shortauthors{Li, Liu et al.}

\title{Millisecond Pulsars in M2: New discoveries and a detailed timing analysis} 

\author[0009-0008-4109-744X]{Baoda Li}
\affiliation{College of Physics, Guizhou University, Guiyang 550025, China}

\author[0000-0002-2953-7376]{Kuo Liu$^{\ast}$}
\affiliation{Shanghai Astronomical Observatory, Chinese Academy of Sciences, 80 Nandan Road, Shanghai 200030, China}
\affiliation{State Key Laboratory of Radio Astronomy and Technology, A20 Datun Road, Chaoyang District, Beijing, 100101, P. R. China}
\affiliation{Max-Planck-Institut f\"ur Radioastronomie, Auf dem H\"ugel 69, D-53121 Bonn, Germany}

\author[0000-0003-0757-3584]{Lin Wang}
\affiliation{Shanghai Astronomical Observatory, Chinese Academy of Sciences, 80 Nandan Road, Shanghai 200030, China}

\author[0009-0001-8816-8523]{Shunyi Lan}
\affiliation{International Centre of Supernovae (ICESUN), Yunnan Key Laboratory of Supernova Research, Yunnan Observatories, Chinese Academy of Sciences, 396 Yangfangwang, Kunming 650216, China}

\author{P. C. C. Freire}
\affiliation{Max-Planck-Institut f\"ur Radioastronomie, Auf dem H\"ugel 69, D-53121 Bonn, Germany}

\author{Pinsong Zhao}
\affiliation{Kavli Institute for Astronomy and Astrophysics, Peking University, Beijing 100871, P.R. China}

\author[0000-0002-2394-9521]{Liyun Zhang$\dag$}
\affiliation{College of Physics, Guizhou University, Guiyang 550025, China}

\author[0000-0002-7909-4171]{Zhengwei Liu}
\affiliation{International Centre of Supernovae (ICESUN), Yunnan Key Laboratory of Supernova Research, Yunnan Observatories, Chinese Academy of Sciences, 396 Yangfangwang, Kunming 650216, China}

\author{Lei Qian}
\affiliation{National Astronomical Observatories, Chinese Academy of Sciences, 20A Datun Road, Chaoyang District, Beijing, 100101, Peopleʼs Republic of China}
\affiliation{Guizhou Radio Astronomical Observatory, Guizhou University, Guiyang 550025, Peopleʼs Republic of China}
\affiliation{100101 Key Laboratory of Radio Astronomy and Technology (Chinese Academy of Sciences), A20 Datun Road, Chaoyang District, Beijing, 100101, P. R. China}
\affiliation{College of Astronomy and Space Sciences, University of Chinese Academy of Sciences, Beijing 100049, Peopleʼs Republic of China}

\author{Wu Jiang}
\affiliation{Shanghai Astronomical Observatory, Chinese Academy of Sciences, 80 Nandan Road, Shanghai 200030, China}
\affiliation{State Key Laboratory of Radio Astronomy and Technology, A20 Datun Road, Chaoyang District, Beijing, 100101, P. R. China}

\author[0000-0001-6051-3420]{Dejiang Yin}
\affiliation{College of Physics, Guizhou University, Guiyang 550025, China}

\author{Yaowei Li}
\affiliation{College of Physics, Guizhou University, Guiyang 550025, China}

\author[0009-0007-6396-7891]{Yinfeng Dai}
\affiliation{College of Physics, Guizhou University, Guiyang 550025, China}

\author{Yang Liu}
\affiliation{Shanghai Astronomical Observatory, Chinese Academy of Sciences, 80 Nandan Road, Shanghai 200030, China}

\author[0000-0001-5316-2298]{Xiangcun Meng}
\affiliation{International Centre of Supernovae (ICESUN), Yunnan Key Laboratory of Supernova Research, Yunnan Observatories, Chinese Academy of Sciences, 396 Yangfangwang, Kunming 650216, China}

\author{Zhichen Pan}
\affiliation{National Astronomical Observatories, Chinese Academy of Sciences, 20A Datun Road, Chaoyang District, Beijing, 100101, China}
\affiliation{Guizhou Radio Astronomical Observatory, Guizhou University, Guiyang 550025, China}
\affiliation{100101 Key Laboratory of Radio Astronomy and Technology (Chinese Academy of Sciences), A20 Datun Road, Chaoyang District, Beijing, 100101, P. R. China}
\affiliation{College of Astronomy and Space Sciences, University of Chinese Academy of Sciences, Beijing 100049, China}

\footnote{$\ast$ liukuo@shao.ac.cn}
\footnote{$\dag$ liy\_zhang@hotmail.com}

\begin{abstract}
Globular clusters (GCs) offer a unique environment for discovering and studying millisecond pulsars. In this paper, we present a multi-epoch search and detailed timing analysis of millisecond pulsars in the GC M2, using the Five-hundred-meter Aperture Spherical Telescope. We have discovered two new binary millisecond pulsars in M2, designated M2F and M2G, respectively. We provide measurements of the emission properties of all known pulsars in M2, including their polarization profiles, rotation measures, flux densities, scintillation characteristics, and so forth. In particular, we report the first rotation measure at the distance and direction of this cluster. Additionally, we report the first phase-coherent timing solutions for the M2 pulsars. From our Bayesian timing analysis, we have measured their spin and orbital parameters with high precision, including the advance of periastron for M2A and M2E indicating total system masses of 1.75(13) and 1.80(5) solar masses respectively. Using archival data from the Hubble Space Telescope, we have identified an optical counterpart of M2C, which is likely the white dwarf companion of the pulsar. By combining results from optical and radio observations, we have reconstructed the binary evolution track of this system and estimated the cooling age of the companion to be approximately 10\,Myr, making it the youngest white dwarf in any known GC binary pulsars. Furthermore, using the spin period derivatives of M2 pulsars, we have investigated the gravitational potential of the cluster and found that our results strongly support the latest central-stellar-velocity dispersion measurement in M2.
\end{abstract}



\section{Introduction} \label{sec:intro}
A Globular cluster (GC) is a spherical collection of stars bound together by gravity. These clusters provide a unique environment where numerous millisecond pulsars (MSPs) can form, driven by their dense stellar populations and frequent dynamical interactions \citep[e.g.,][]{ver87,vf14}.
Since the first pulsar discovered in a GC \citep{lbm+87}, a total of 345 pulsars have been found in 45 GCs\footnote{\url{https://www3.mpifr-bonn.mpg.de/staff/pfreire/GCpsr.html}}. These discoveries include several peculiar pulsar systems, such as the fastest spinning pulsar known to date \citep{hrs+06}, a triple pulsar system \citep{tacl99}, a transitional X-ray binary \citep{pfb+13}, eccentric binary MSPs with massive companions \citep[e.g.,][]{fgri04,bdf+24}, the binary pulsar with the shortest known orbital period \citep{plj+23}, and a binary pulsar with a compact companion in the mass gap between neutron stars (NSs) and black holes \citep{bdf+24}.

GC MSPs serve as exceptional astrophysical laboratories for a variety of studies. Firstly, timing of binary MSPs is an excellent tool for testing gravity theories in the strong-field regime \citep{ksm+21,fw24} and probing the equation of state of nuclear matter above the density of the atomic nucleus \citep[e.g.,][]{of16}, especially by finding very massive NSs \citep{afw+13,fcp+21}. These investigations are typically achieved by precisely measuring the orbital parameters of binary systems using the pulsar timing technique. This works by measuring the Keplerian parameters first, and then start to measure the post-Keplerian parameters as the observing baseline extends. 
The post-Keplerian parameters quantify relativistic effects in both the orbital dynamics and the propagation of light within the system.
Assuming a theory of gravity, we can use them to derive the masses of the two objects and to test the theory of gravity itself.

In parallel, precise localization of the pulsars from the timing analysis will allow for multi-wavelength follow-up of the system, helping to determine the nature of the binary \citep[e.g.,][]{ccp+20,lfc+25}. For bright white-dwarf companions, optical observations may allow for effective temperature and radial velocity measurements, yielding mass measurements and evolution track of the binary system \citep{afw+13,cfi+19,bbh+21}.

Additionally, finding a MSP--black hole system, which is anticipated to form only in an environment with a high stellar density like the center of a GC, will offer an unprecedented opportunity to test some fundamental pillars of black hole physics, such as the cosmic censorship conjecture and the no-hair theorem \citep{wk99,lwk+12,lewk14}. Such a system might have been recently discovered \citep{bdf+24}.

In addition, finding a collection of MSPs in a single GC allows unique studies of the cluster itself.
Timing analyses can yield precise positions, period derivatives ($\dot{P}$), and proper motions of GC pulsars. 
Since the measured variation in their spin periods is primarily due to the acceleration within the cluster itself, these measurements can be used to map the cluster's gravitational potential \citep[e.g.][]{frk+17,prf+17,apr+18}. 
If a pulsar is located close to the cluster core, measurements of its higher order period derivatives may become achievable. This could be used to investigate the mass concentration in the center of the cluster, where an intermediate-mass black hole may inhabit \citep[e.g.,][]{psl+17} or to confirm the detection of non-luminous matter \citep{Abb+19}.

Radio observations also provide precise measurements of the rotation measures (RMs) of the pulsars in a single GC, offering insights into the cluster's magnetic field. If an RM gradient for pulsars within the same cluster is detected, this then provides the opportunity to search for a potential magnetized outflow originated from the Galactic plane \citep{apt+20,mrd+22}. Moreover, studying a group of pulsars in a GC will allow to track the dynamical evolution of the cluster \citep[e.g.,][]{pdm+03} and studying the interstellar medium (ISM) along different lines of sight (LoS) through the cluster \citep[e.g.,][]{fkl+01,apr+18,lzy+24}.

In this paper, we present an in-depth study of pulsars in the GC M2, using the Five-hundred-meter Aperture Spherical radio Telescope \citep[FAST,][]{nlj+11}. 
M2 (NGC 7089) is approximately 11.69\,kpc away from the Earth, its metallicity ([Fe/H]) is -1.65 and has a escape velocity of 43.6 $\rm {km/s}$ \citep{bv21}. 
M2 is one of the GCs in the FAST sky. With its highly sensitive searches, FAST has enabled the discovery of the first five pulsars in M2, namely M2A to M2E, all of which are binary millisecond pulsars \citep{pqm+21}. The spin periods of these pulsars range from 3 to 11\,ms, and their dispersion measures (DMs) are approximately between 43.3 and 44.1\,pc\,cm$^{-3}$. The orbital periods of M2A to M2D span from 1 to 10\,days, while an initial timing solution for M2E could not be obtained due to the limited number of detections. Based on the estimation by \citet{yzq+24}, there could be at least 80 detectable pulsars in M2 using FAST, which motivates ongoing search efforts in this cluster.

The structure of the paper is as follows. Section~\ref{sec:obs} describes the observations and data processing methods used in this work. Section~\ref{sec:res} presents the discoveries of the pulsar search, the emission properties and timing analysis of all known M2 pulsars. Section~\ref{sec:dis} discusses the properties of the M2 pulsars and the cluster, based on the observational results from radio and other frequency bands. Finally, Section~\ref{sec:conc} summarizes the key findings of this paper.

\section{Observations and data reduction} \label{sec:obs}
\subsection{FAST observations of GC M2}
FAST conducted observations of M2 on six epochs from MJDs 58795 to 58852, primarily to search for new pulsars and obtain initial timing solutions for newly discovered pulsars in this cluster \citep{pqm+21}. Since MJD 59483, a regular monitoring campaign has been underway to perform precision timing analysis of M2 pulsars while continuing the search for new pulsars. In total, 31 observations from MJDs 58795 to 60512 were analyzed for this work. During all observations, the telescope was pointed toward the center of M2 at Right Ascension (RA) 21:33:27.02 and Declination (DEC) $-$00:49:23.7 in the J2000 coordinate system \citep{bv21}, with exposure times ranging approximately from 1 to 2\,hr (see Table~\ref{tab:obs}). The radio signal was collected using the 19-beam L-band receiver, which covers a frequency range from 1.0 to 1.5\,GHz. Since the core radius of M2 ($0.32'$) fits well within the Full Width at Half Maximum of a single receiver ($3'$), only data from the central beam were recorded. The data were sampled in 8-bit resolution and recorded in \textsc{psrfits} search-mode format, with a sampling time of 49.152\,$\mu$s and 4096 frequency channels (corresponding to a 0.122-MHz channel width). Data were recorded in dual polarization for observations before MJD 59662. For each observation afterward, an off-source scan with a noise diode injected into the receiver was conducted prior to the exposure on M2 for polarization calibration. Consequently, the data were recorded with the full Stokes polarization product. Additionally, on MJDs 59763 and 59787, a total of 10-min on/off scans of 3C394 were conducted to calibrate the flux density of the epoch observations.

\subsection{Data processing} \label{ssec:data_proc}

We employed the \texttt{PRESTO} software package to perform blind searches for new pulsars in M2 \citep{rem02}. 
In order to catch possible pulsars with short orbital period, we searched with the whole length of the data and about 30 minutes segments.
The routine {\tt rfifind} was utilized to generate radio frequency interference (RFI) masks. The length of data for finding radio-frequency-domain RFI was 2.0\,s. The routine {\tt prepdata} was then applied to correct for dispersion in the radio signal data. The DM range used was 38--48\,pc\,cm$^{-3}$, with a step size of 0.01\,pc\,cm$^{-3}$, considering a possible DM range in a given GC \citep{yzl+23}. After applying a Fast Fourier Transform to the time-series data, the routine \texttt{accelsearch} was used to search for periodic signals in the Fourier spectrum. The {\tt -zmax} value, representing the maximum number of Fourier frequency bins that a signal can drift during an individual observation due to acceleration of the pulsar, was set up to be 1200 in the searches. All identified periodic signals from a single DM trial were folded using the routine {\tt prepfold} to produce pulsar candidates which were subsequently visually inspected. These analyses led to the discovery of two new millisecond pulsars, namely M2F and M2G (see Section~\ref{ssec:M2FG}).

To obtain the initial timing solution of M2 pulsars, we first searched for these pulsars in all epoch observations and used the \texttt{prepfold} routine to fold the data individually to collect a number of pulse-profile detections. Based on those detections, for each pulsar we utilized the {\tt fitorb.py} routine to determine its spin period and the Keplerian parameters of its binary orbit. It is worth noting that while initial timing solutions for M2A to M2D previously reported in \citet{pqm+21} were obtained in the same way, here they have been refined using a significantly larger number of epoch observations.

To establish the first phase-coherent timing solutions for the M2 pulsars, we first used the initial timing solutions mentioned above to fold all epoch observations. Then, for each pulsar, we employed the {\tt get\_TOAs.py} routine in \texttt{PRESTO} to measure the pulse time of arrivals (TOAs). The template profile used for the TOA measurement was created by fitting the pulse profiles with the highest signal-to-noise ratios (S/Ns) using the {\tt pygaussfit.py} routine. Next, we employed the \textsc{dracula}\footnote{\url{https://github.com/pfreire163/Dracula}}, \texttt{TEMPO}\footnote{\url{https://tempo.sourceforge.net/}} and \texttt{TEMPOTWO} \citep{hem05a} software packages to connect the TOAs in phase and obtain the first phase-coherent timing solution, by following the methods described in \citet{fr18}. 

To refine the timing analysis results, we folded all epoch observations using these first phase-coherent timing solutions with the \texttt{DSPSR} software package to create archival data of pulse profiles, which were further reduced using a collection of routines from the \texttt{PSRCHIVE} software package. Data collected after MJD 59662 were polarization-calibrated with the {\tt pac} routine using the \textsc{SingleAxis} method, using noise diode scans conducted prior to each epoch observation. This process corrected for differential phase and amplitude between the two polarizations. For the data from MJDs~597663 and 59787, we also calibrated the flux density using the scans of 3C394 conducted on the same day. The data were cleaned of narrow-band RFIs using both an automatic running median zapping method in the {\tt paz} routine and visual inspections with the {\tt pazi} routine. For each epoch observation, we averaged the data in time and frequency to create two sub-integrations, each with four 125-MHz frequency bands, except for M2F and M2G, where we created only one integration due to the low peak S/Ns of the detections (see Table~\ref{tab:obs}). We then measured TOAs from the profiles in each integration and frequency band using the canonical template-matching scheme built in the \texttt{pat} routine \citep{tay92}. The templates were created from the epoch observations with the highest S/N for each pulsar, using a wavelet smoothing method implemented in the {\tt psrsmooth} routine. Finally, we refitted the timing solution of each pulsar with the new set of TOAs. To subtract the motion of the Earth with respect to the solar-system barycenter from our timing, we used the Jet Propulsion Laboratory's DE440 Solar System ephemeris \citep{pfwb21}. We employed TT(BIPM2022) as the clock reference, and TDB units for the timing parameter fit. We also attempted to incorporate additional parameters into the timing model of each pulsar, such as proper motion of the pulsar and post-Keplerian parameters of the orbit, retaining those measured with high significance (2--3$\sigma$), following similar steps outlined in \citet{EPTA+2023a}.

\begin{table*}[]
    \begin{tabular}{cccccccccc}
\hline
Date & MJD & $T_{\rm obs}$ (min) & M2A & M2B & M2C & M2D & M2E & M2F & M2G \\
\hline
20191108 & 58795.5 & 63.1 & 13.4 & 8.7 & 17.3 & 9.1 & 6.6 & 5.5 & 7.3 \\  
20191116 & 58803.4 & 120.0 & 13.8 & 14.3 & 22.8 & 13.0 & 14.0 & 7.0 & 5.4 \\  
20191117 & 58804.4 & 119.5 & 13.4 & 15.6 & 17.6 & 15.7 & 8.4 & 7.3 & 5.1 \\  
20191118 & 58805.4 & 114.4 & 11.4 & 12.7 & 15.3 & 8.4 & 9.8 & 7.5 & 6.0 \\  
20191127 & 58814.4 & 60.0 & 11.3 & 15.9 & 15.1 & 8.9 & 5.9 & 6.8 & 6.5 \\  
20200104 & 58852.3 & 120.0 & 20.0 & 14.4 & 26.5 & 8.9 & 13.3 & 6.2 & 9.5 \\  
20210926 & 59483.5 & 119.8 & 10.5 & 18.1 & 25.2 & 22.0 & 16.3 & 6.4 & 6.9 \\  
20210927 & 59484.5 & 119.8 & 8.4 & 15.5 & 17.6 & 18.3 & 19.5 & 6.8 & 5.6 \\  
20210929 & 59486.5 & 119.8 & 5.6 & 15.6 & 30.1 & 30.4 & 12.6 & 7.2 & 8.4 \\  
20211003 & 59490.5 & 119.8 & 8.7 & 17.2 & 13.0 & 10.4 & 9.7 & 7.3 & 7.7 \\  
20211011 & 59498.5 & 119.8 & 9.3 & 14.1 & 27.8 & 14.8 & 6.5 & 8.1 & 6.9 \\  
20211027 & 59514.5 & 119.3 & 9.1 & 10.4 & 31.5 & 19.3 & 14.9 & 6.4 & 8.9 \\  
20211207 & 59555.3 & 119.8 & 13.3 & 16.6 & 41.1 & 18.0 & 12.3 & 5.2 & 5.8 \\  
20211220 & 59568.3 & 119.8 & 13.0 & 11.4 & 21.1 & 17.5 & 15.3 & 6.5 & 15.3 \\  
20220116 & 59595.3 & 119.8 & 16.4 & 22.0 & 32.8 & 13.4 & 9.5 & 9.1 & 8.7 \\  
20220125 & 59604.2 & 119.8 & 15.7 & 22.4 & 32.8 & 15.3 & 7.0 & 4.1 & 9.1 \\  
20220324 & 59662.1 & 123.3 & 19.4 & 14.9 & 45.2 & 21.3 & 12.3 & 6.5 & 11.6 \\  
20220410 & 59679.0 & 123.3 & 15.8 & 14.0 & 39.5 & 27.4 & 18.1 & 9.8 & 7.6 \\  
20220509 & 59708.0 & 123.3 & 11.4 & 13.1 & 24.2 & 18.3 & 15.7 & 5.3 & 6.5 \\  
20220612 & 59741.9 & 88.8 & 11.6 & 12.7 & 37.8 & 11.5 & 8.8 & 6.1 & 7.9 \\  
20220704 & 59763.8 & 133.3 & 18.6 & 16.3 & 38.7 & 10.9 & 6.3 & 5.9 & 12.5 \\  
20220728 & 59787.7 & 133.1 & 17.9 & 16.4 & 19.6 & 9.1 & 12.8 & 6.2 & 5.9 \\  
20221003 & 59855.5 & 119.2 & 11.3 & 12.6 & 25.2 & 12.9 & 15.3 & 9.0 & 8.0 \\  
20221224 & 59937.3 & 119.2 & 7.5 & 8.3 & 11.9 & 12.8 & 8.8 & 5.9 & 6.1 \\  
20230207 & 59982.2 & 103.0 & 8.7 & 6.8 & 12.8 & 15.8 & 5.1 & 6.5 & 5.6 \\  
20230410 & 60044.0 & 119.2 & 11.5 & 12.8 & 22.6 & 8.8 & 10.8 & 5.6 & 16.6 \\  
20230616 & 60110.8 & 119.2 & 10.6 & 10.3 & 20.4 & 8.6 & 7.4 & 5.7 & 8.1 \\  
20230701 & 60125.8 & 119.2 & 12.7 & 12.2 & 22.1 & 6.3 & 7.0 & 5.3 & 6.1 \\  
20240315 & 60384.1 & 115.1 & 13.5 & 12.8 & 9.7 & 16.0 & 5.1 & 4.8 & 6.2 \\  
20240515 & 60444.9 & 115.1 & 8.8 & 10.5 & 19.6 & 9.3 & 6.5 & 7.1 & 4.9 \\  
20240721 & 60511.7 & 115.2 & 8.5 & 7.8 & 38.5 & 10.7 & 9.3 & 6.6 & 10.5 \\
\hline
    \end{tabular}
    \caption{ List of FAST observations of M2 used for this paper. From the left to right, the columns give the date, MJD, length of the observation, and the peak S/N of detection for the M2 pulsars. }
    \label{tab:obs}
\end{table*}

\section{Results} \label{sec:res}
\subsection{Discovery of M2F and M2G} \label{ssec:M2FG}

\begin{figure}[h]
\centering
\includegraphics[width=\columnwidth]{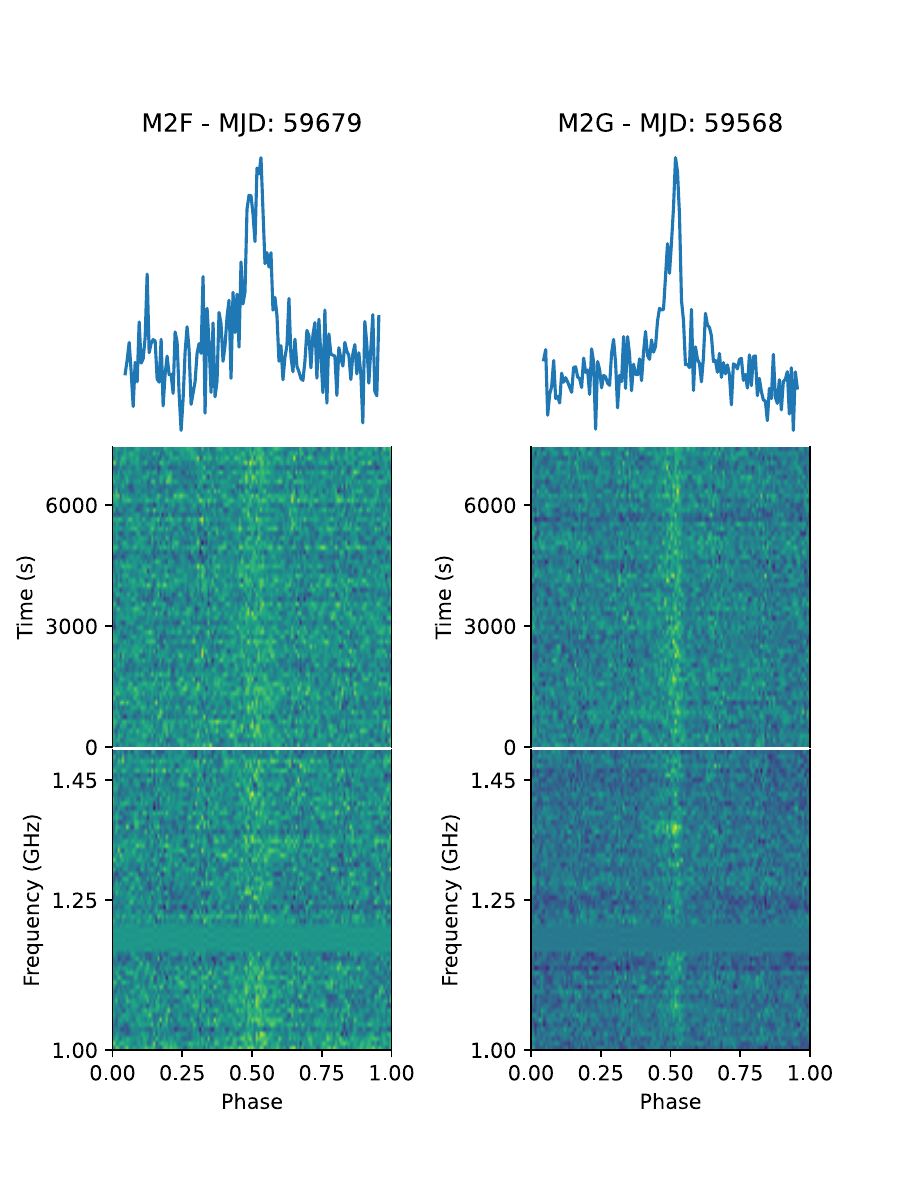}
\caption{Average pulse profiles as well as its time-domain and frequency-domain detections of M2F (from MJD 59679) and M2G (from MJD 59568).}
\label{fig:M2FG}
\end{figure}

Two new binary millisecond pulsars, namely M2F and M2G, were discovered in our searches for pulsars in M2 (see Figure \ref{fig:M2FG} for the first detected signal). The spin period and DM at the time of discovery were 4.78\,ms, 43.40\,pc\,cm$^{-3}$ for M2F, and 2.54\,ms, 43.30\,pc\,cm$^{-3}$ for M2G. Given that the DMs of M2F and M2G are very close to those of the previously discovered pulsars M2A to M2E, it is highly likely that these new pulsars are associated with the M2 cluster. These two new pulsars are extremely faint and their flux densities as well as probability to be detected are also strongly affected by interstellar scintillation (see Section~\ref{ssec:emission} for more details). The discoveries were made with new observations by FAST after \citet{pqm+21}.
In detail, M2F was first detected with data collected on MJD~59679 and M2G was discovered using data from MJD 59763. From a small number of epoch detections, both new pulsars exhibited spin period changes, suggesting that they are in a binary system.

\subsection{Emission properties of the M2 pulsars}  \label{ssec:emission}

\begin{figure*}
\hspace*{-1.0cm}
\includegraphics[scale=0.27]{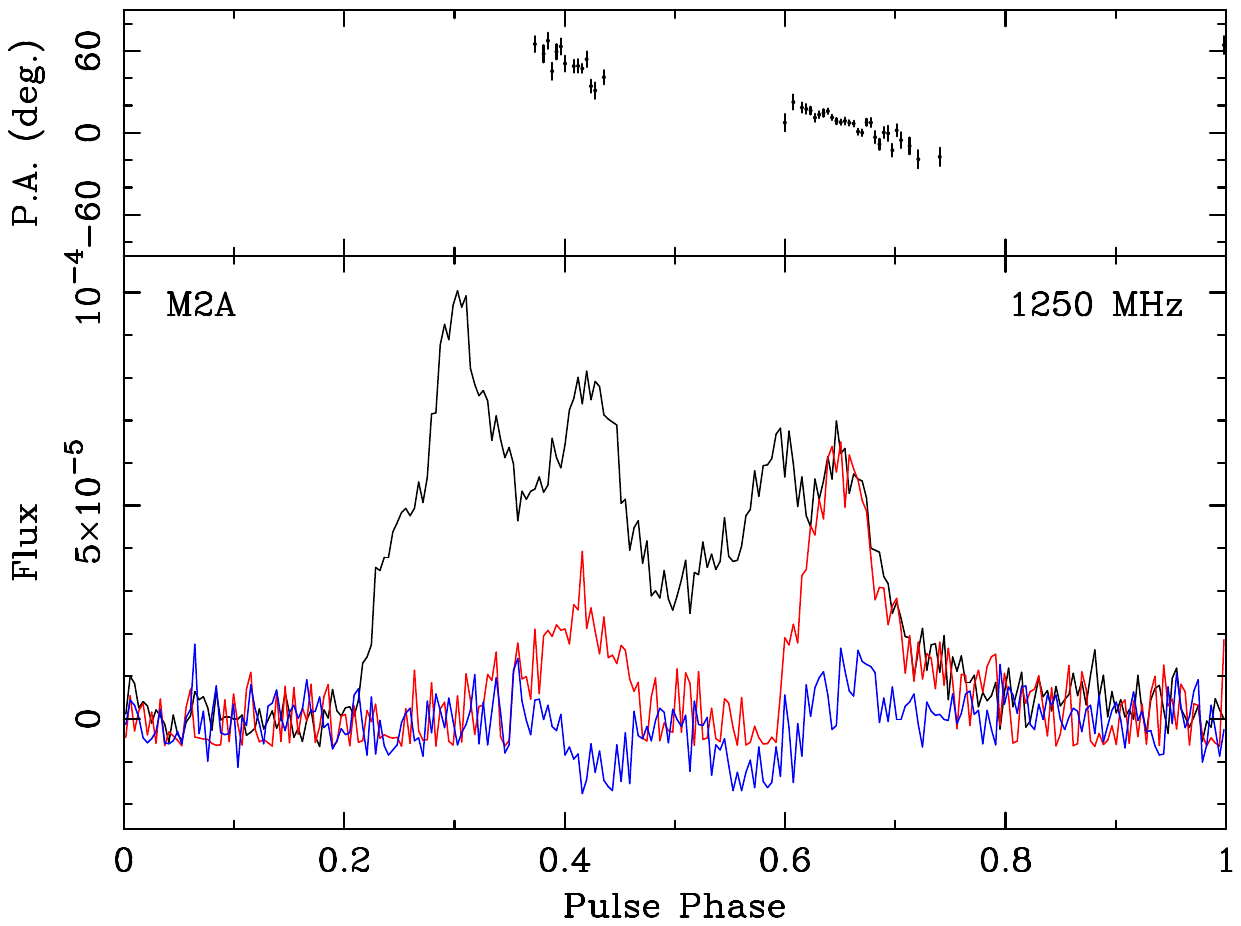}
\hspace*{-1.7cm}
\includegraphics[scale=0.27]{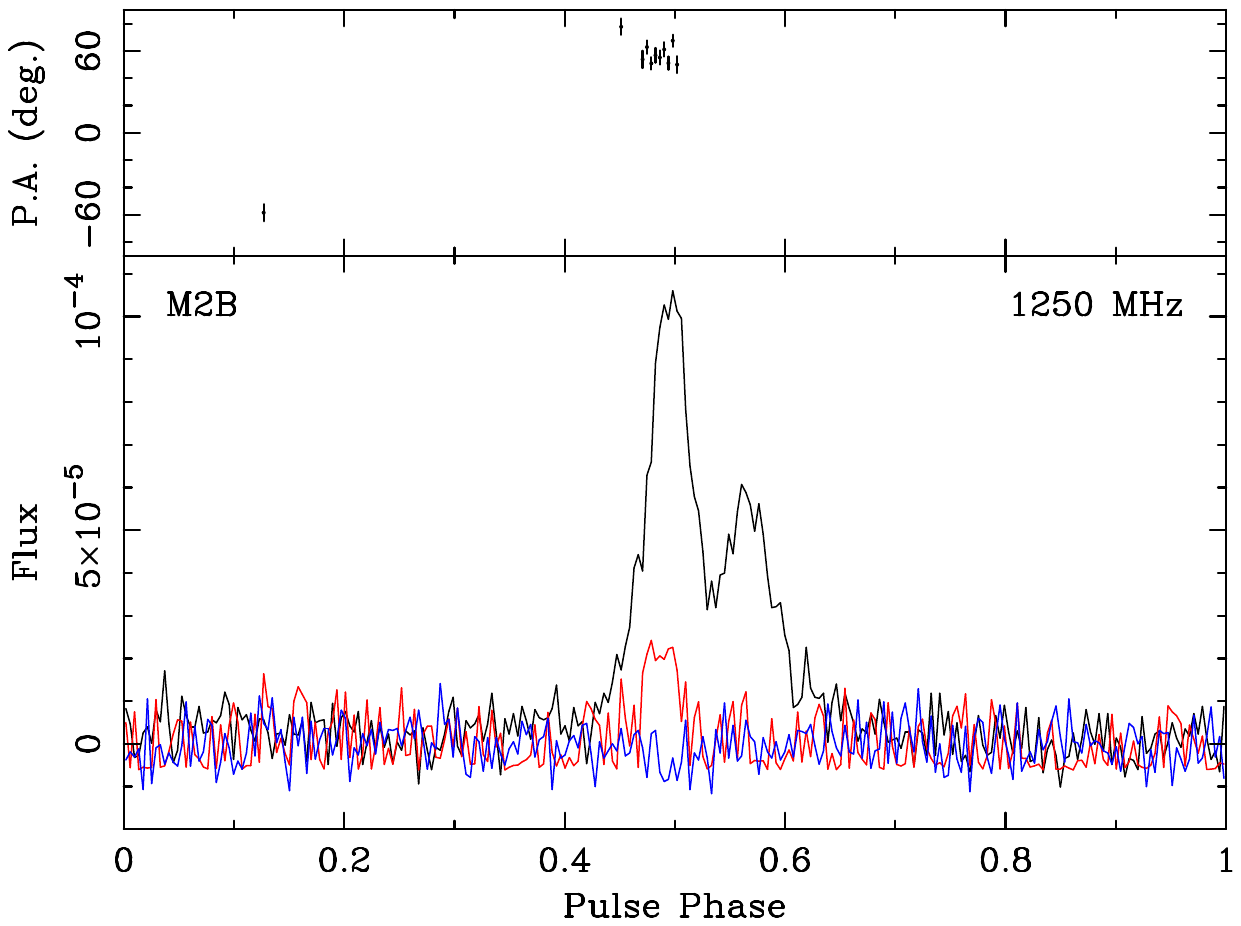}
\hspace*{-1.6cm}
\includegraphics[scale=0.27]{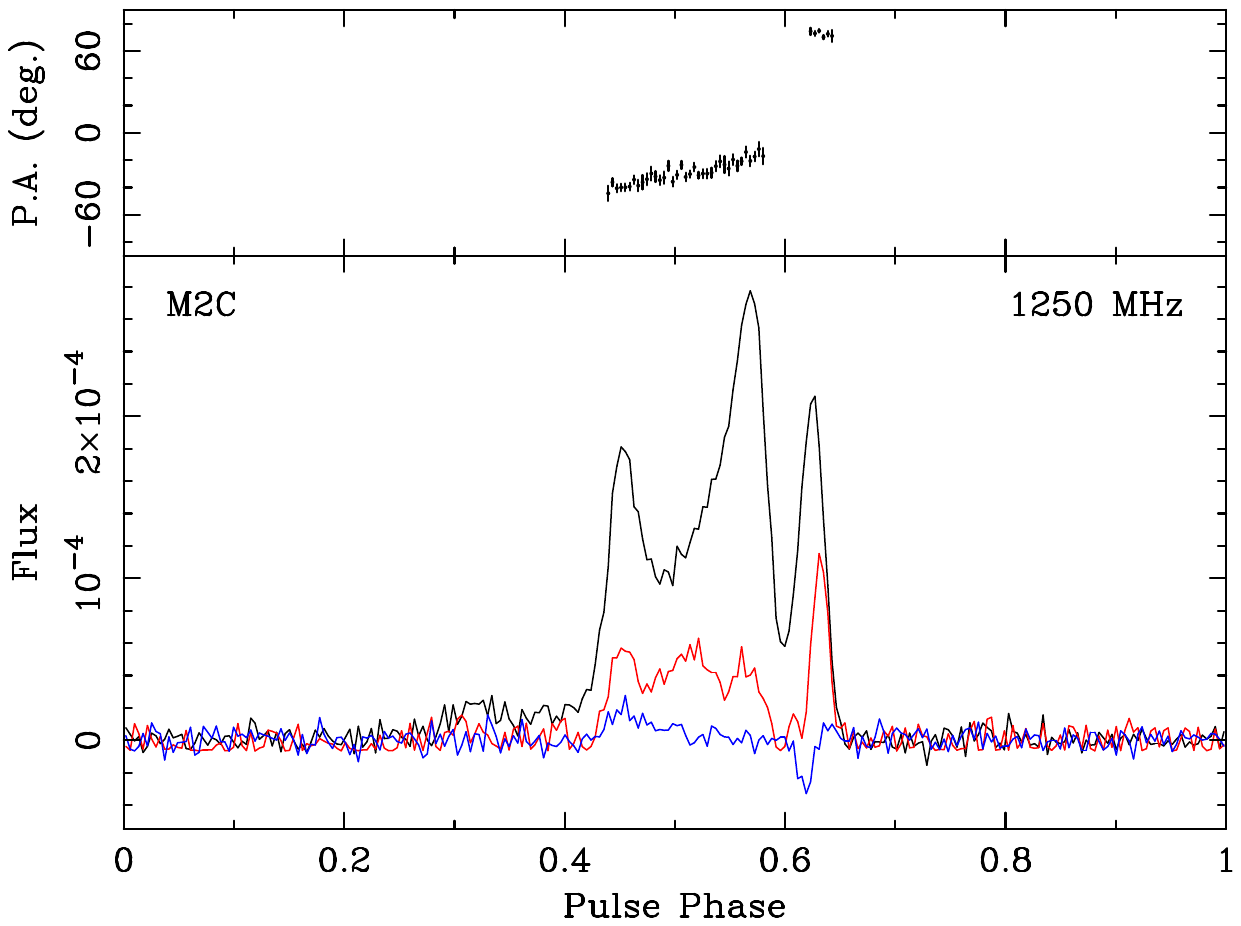}

\vspace*{-1cm}

\hspace*{-1.0cm}
\includegraphics[scale=0.27]{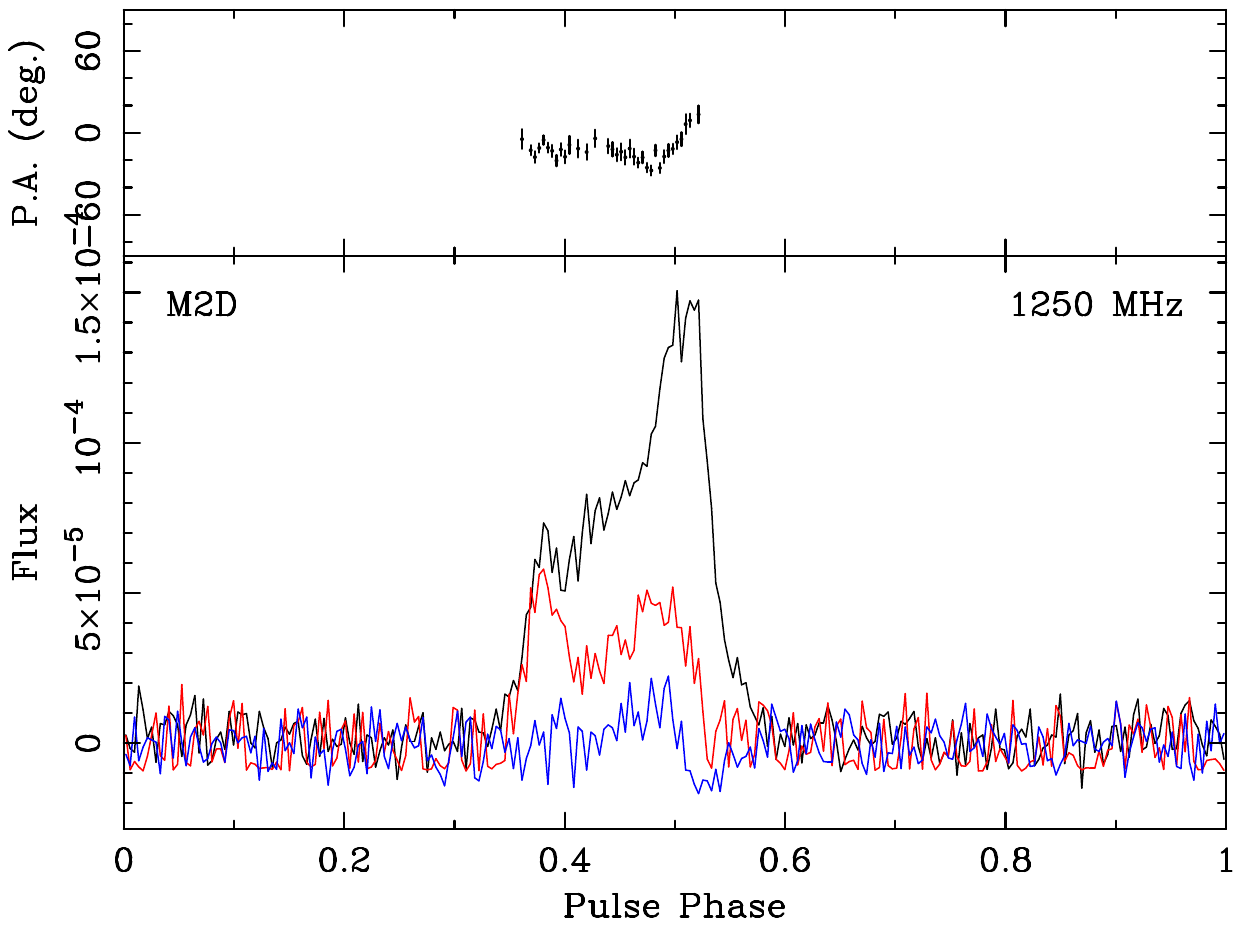}
\hspace*{-1.7cm}
\includegraphics[scale=0.27]{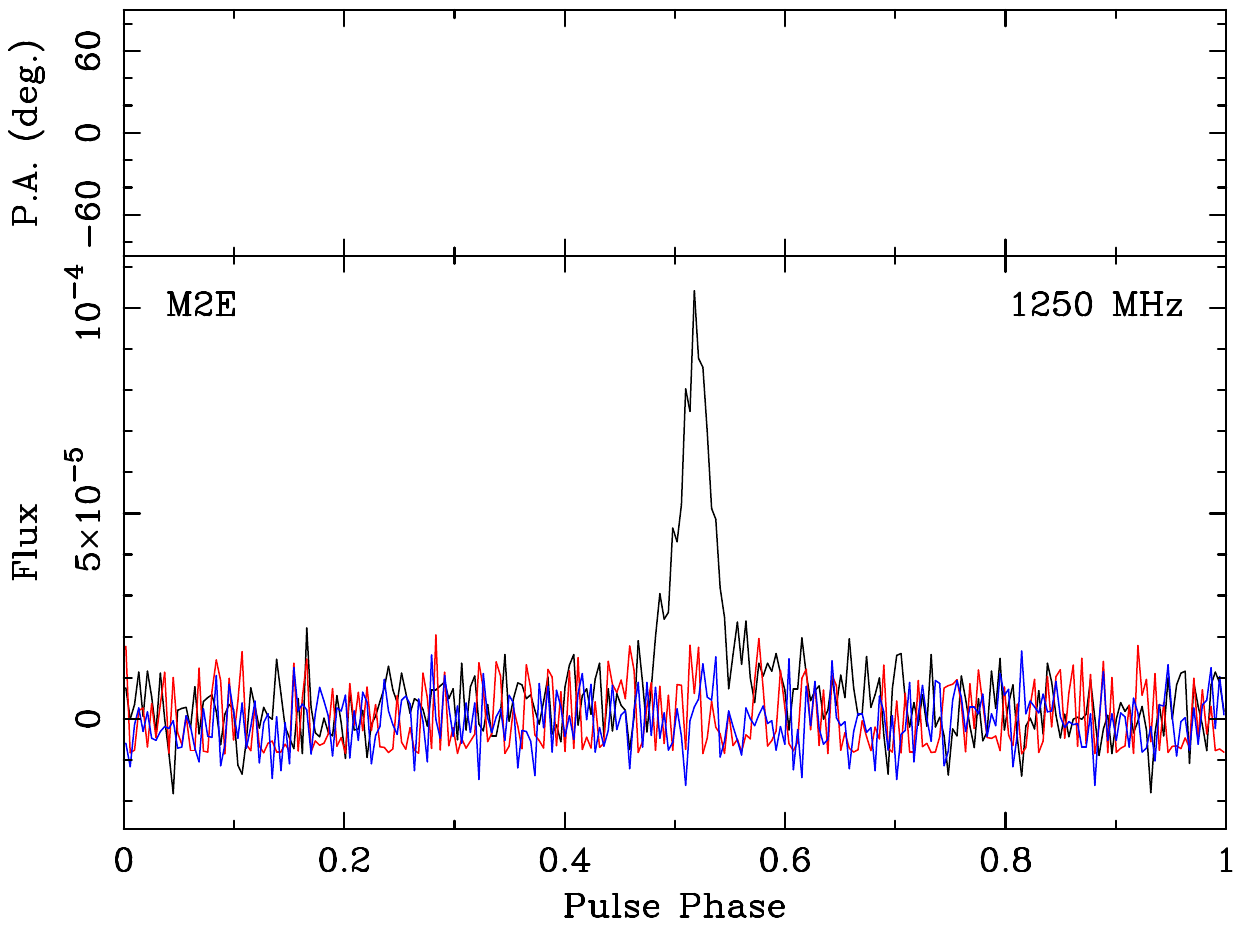}
\hspace*{-1.6cm}
\includegraphics[scale=0.27]{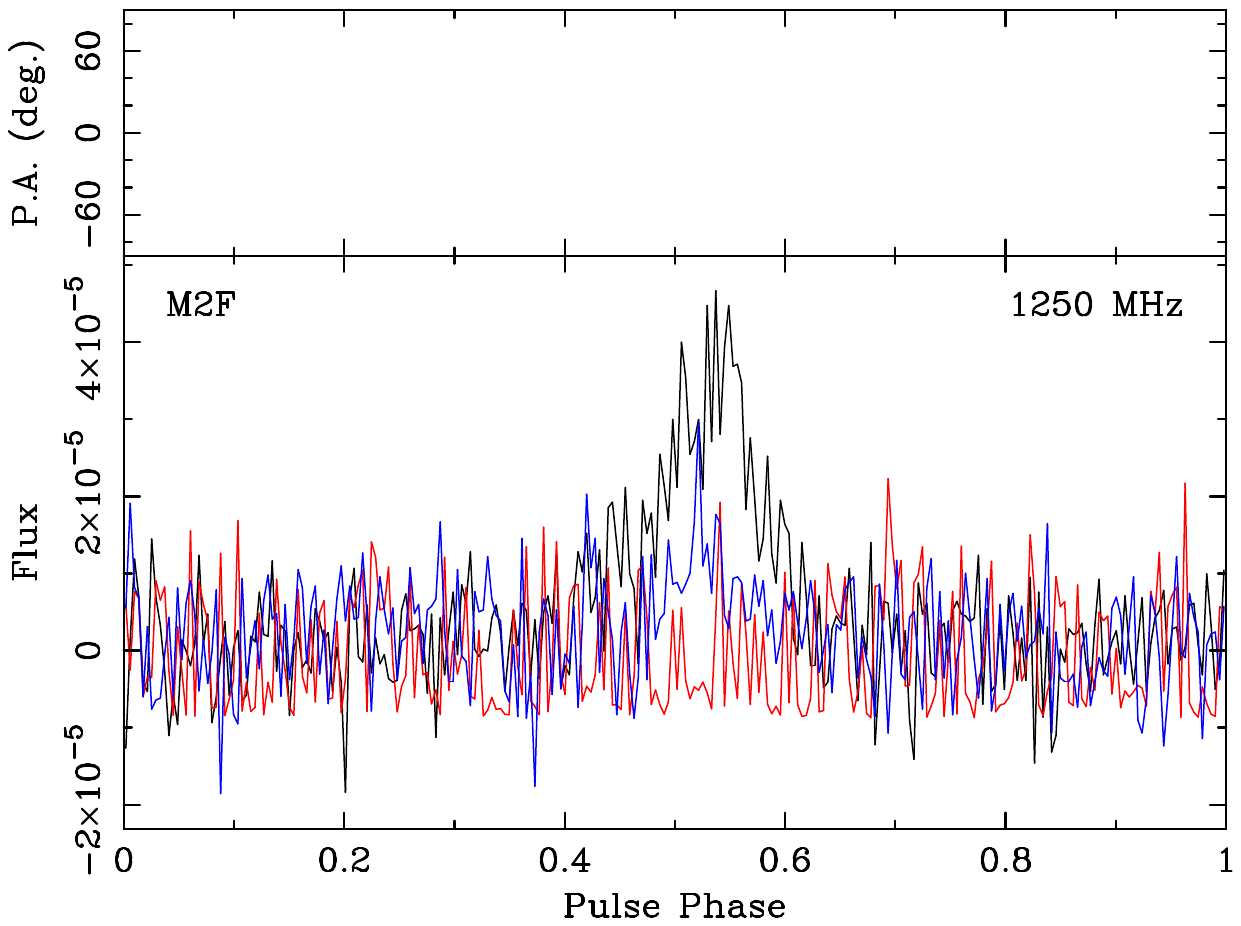}

\vspace*{-1cm}

\hspace*{-1.0cm}
\includegraphics[scale=0.27]{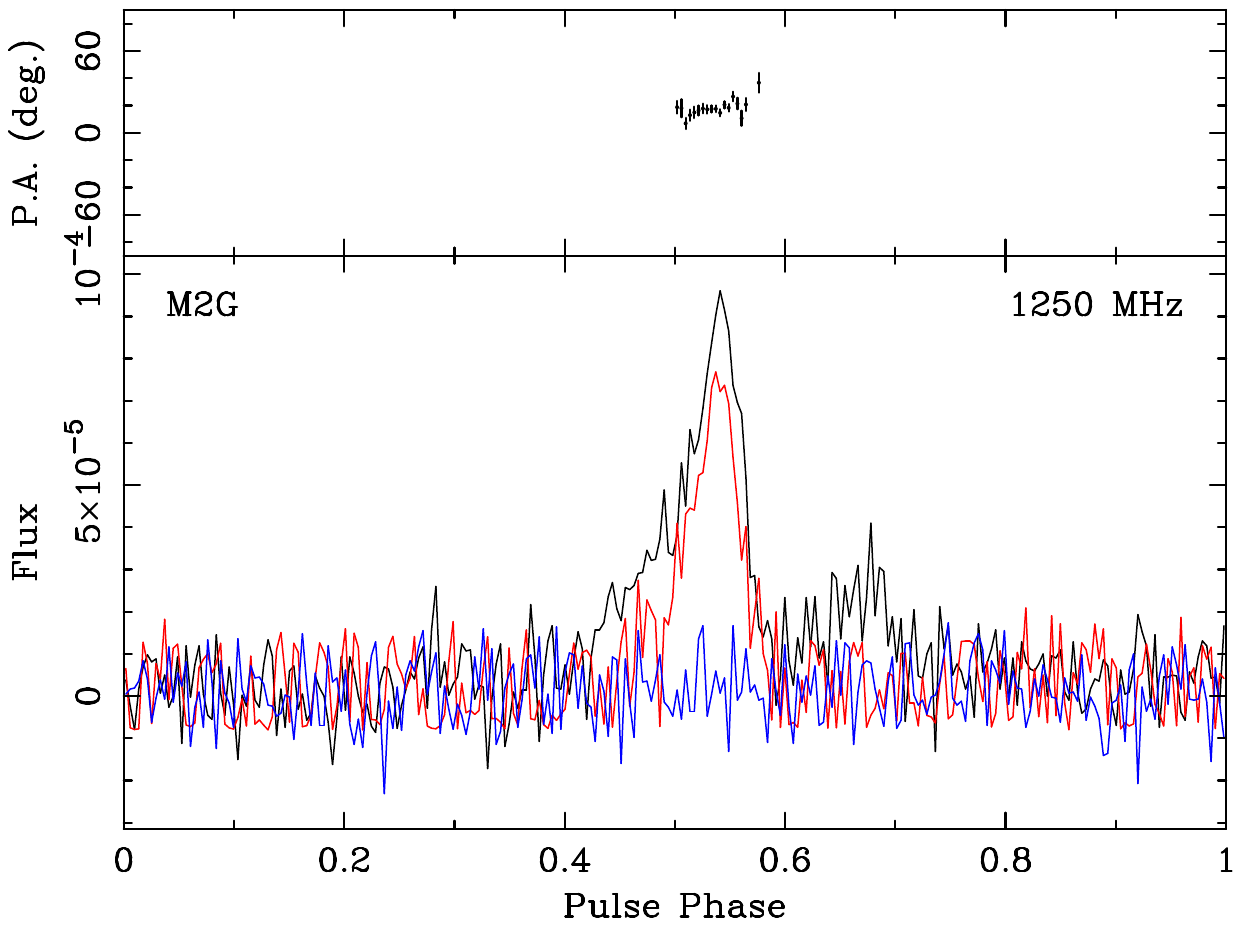}
\caption{Polarization profiles of the M2 pulsars. The black, red, and blue lines represent total intensity ($I$), linear ($L$), and circular ($V$) components, respectively. In each subplot, the top panel shows the linear polarization position angles. For M2A, M2C, M2D, and M2G, the measured RM values presented in Table~\ref{tab:prof_mes} were utilized to corrected for the Faraday rotation in the polarization profiles. For the rest of the pulsars, a weighted mean of these values (25.4\,rad/m$^{2}$) were used instead.
\label{fig:polprof}}
\end{figure*}

Using epoch observations with full Stokes detections, we constructed high S/N integrated profiles for each pulsar using the ephemerides from the timing analysis (see Section~\ref{ssec:timing}). Figure~\ref{fig:polprof} displays the polarization profiles of the M2 pulsars. Significant polarized emission was detected in M2A, M2B, M2C, M2D, and M2G. Based on these polarization profiles, we were able to detect significant Faraday rotation and determine RMs for four of the M2 pulsars: M2A, M2C, M2D, and M2G. The RMs were obtained using the \texttt{rmfit} routine in the \texttt{PSRCHIVE} software pacakge. As shown in Table~\ref{tab:prof_mes}, the RMs from different pulsars are consistent with each other, yielding a weighted mean of $25.4\pm4.0$\,rad/m$^{2}$ for the cluster's RM. The measurement precision still needs improvement in order to investigate whether an RM gradient exists within these pulsars in the cluster as observed for the pulsars in Terzan 5 \citep[see Figure 2 of][]{mrd+22}.

\begin{figure}
\hspace{-1.6cm}
\includegraphics[scale=0.55]{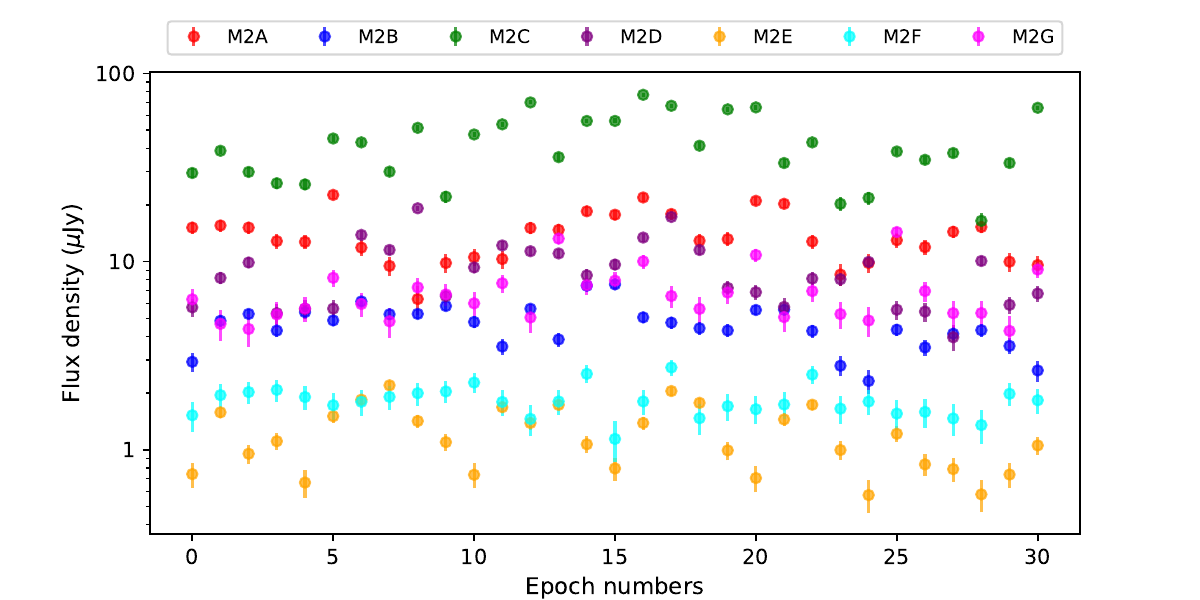}
\caption{Flux density variation of the M2 pulsars from different epoch observations. }
\label{fig:Flux}
\end{figure}

The pulse widths of the M2 pulsars are generally among the typical values for MSPs (0.1--1\,ms). For each pulsar, we derived its flux densities across all epoch observations. The measurements from MJD~59763 and 59787 were obtained from flux calibration, and the rest were based on their relative S/N (per unit integration time) to those of these two epochs. From Figure~\ref{fig:Flux}, it can be seen that the flux densities of the M2 pulsars range from 1 to 50\,$\mu$Jy, with M2C being the brightest and M2E the faintest. For each pulsar, the flux density also exhibits significant variation across different epochs. This variation is typically the result of diffractive scintillation caused by the interstellar medium along the LoS of the pulsar, and a more detailed study of such is presented in Section~\ref{ssec:scint}. 

\begin{table*}
\centering
\begin{tabular}{ccccccc}
\hline
Name & L/I (\%) & $|V|$/I (\%) & RM (rad/m$^{2}$) & W$_{\rm 50}$ (ms)  & W$_{\rm 10}$ (ms) & S ($\mu$Jy) \\
\hline
M2A & 29.9(5) & 9.0(3) & 26.6(9) & 0.743 & 3.32 & 13\\
M2B & 14.7(9) & --- & ---  & 0.298 & 1.02 & 4.7 \\
M2C & 27.3(4) & --- & 25.8(6) & 0.209 & 0.694 & 39\\
M2D & 39(1) & 3.3(4) & 22.9(1.1)  & 0.180 & 0.604 & 8.2\\
M2E & --- & --- & ---  & 0.107 & 0.260 & 1.1 \\
M2F & --- & --- &  ---  & 0.121 & 1.05  & 1.8 \\
M2G & 53.0(2) & --- & 24.1(1.3)  & 0.0859 & 0.378 & 6.3\\
\hline
\end{tabular}
\caption{Emission properties of the M2 pulsars. From the left to right the columns are pulsar name, fractional linear ($L$), and circular ($V$) polarization, rotation measure, width at 50\% and 10\% peak of the pulse profile, and flux density, respectively. The bias in polarization L and $|$V$|$ has been corrected by following schemes described in \cite{wk74} and \cite{kj04}. Note that due to limited by S/N, in some cases a significant measurement cannot be achieved. The flux density of each pulsar here is the median of the measurements shown in Figure~\ref{fig:Flux}. \label{tab:prof_mes}}
\end{table*}

\subsection{Timing analysis of the M2 pulsars} \label{ssec:timing}

\begin{figure*}
\hspace*{-1.1cm}
\includegraphics[scale=0.8]{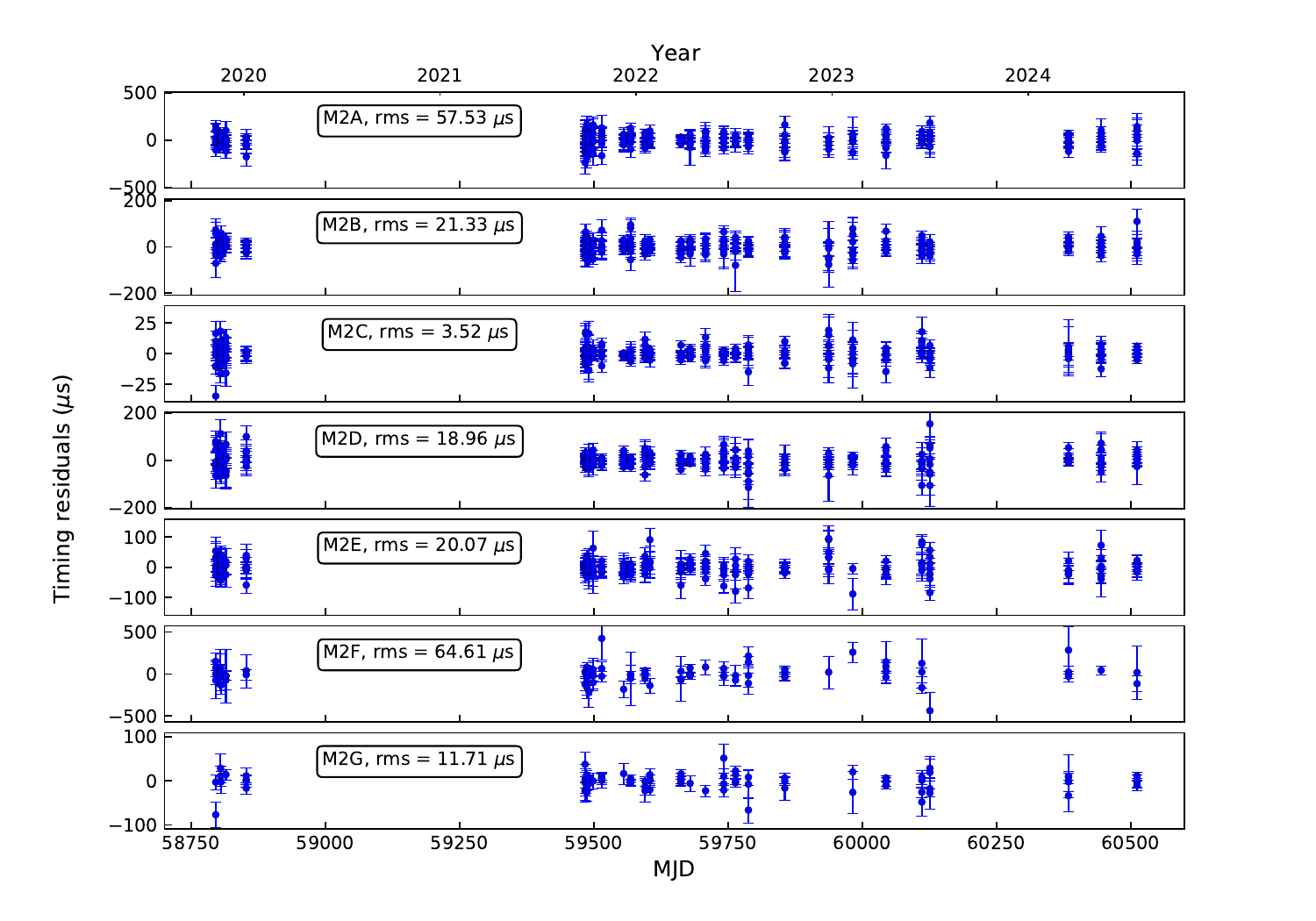}
\caption{ Timing residuals as a function of time for the M2 pulsars, obtained with the timing solutions in Tables~\ref{tab:ephem1} and \ref{tab:ephem2}. Here the white noise parameters have been taken into account in the TOA errors. }
\label{fig:timing_res}
\end{figure*}

\begin{figure*}
\hspace*{-1.1cm}
\includegraphics[scale=0.8]{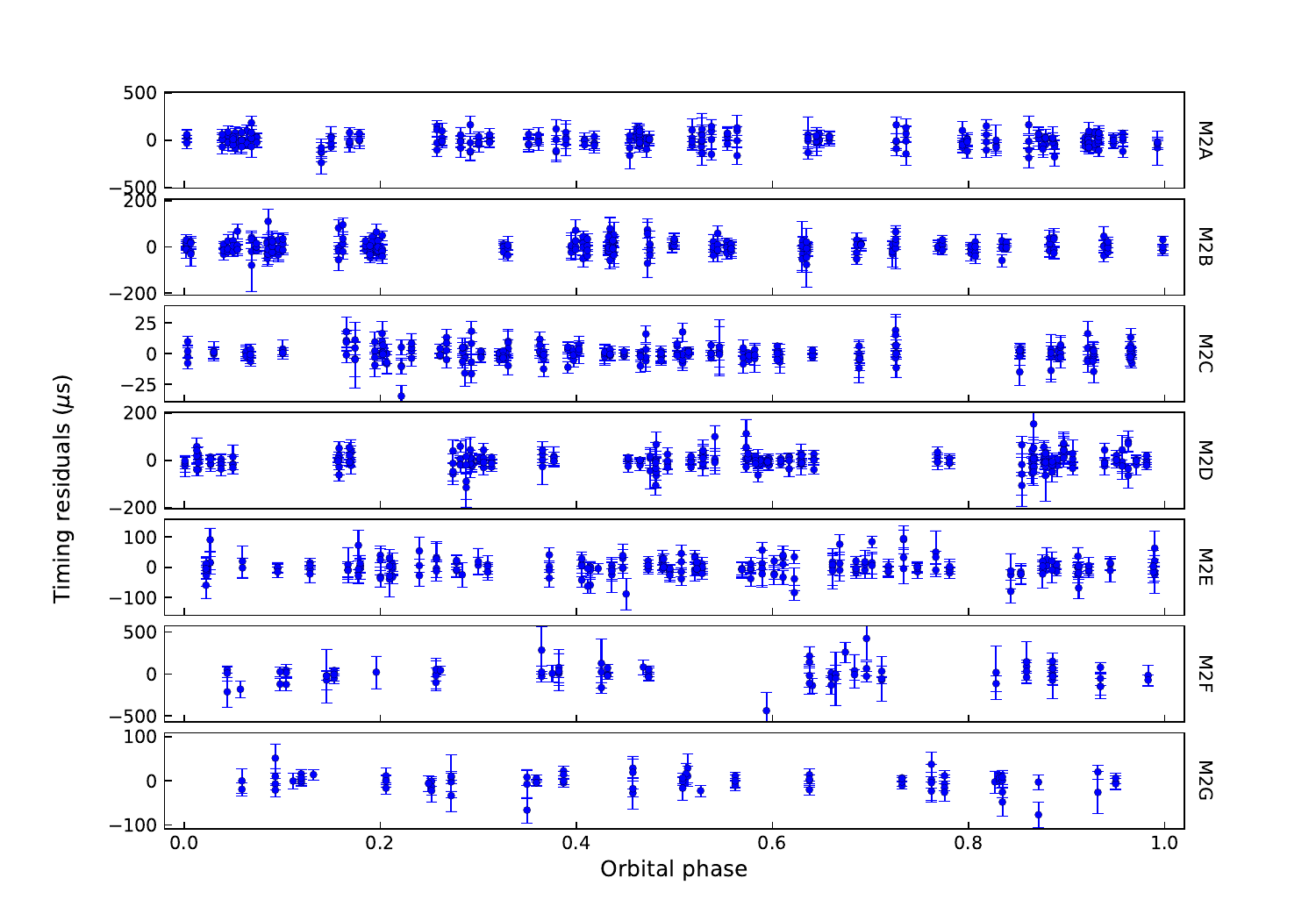}
\caption{ Timing residuals as a function of orbital phase for the M2 pulsars, obtained with the timing solutions in Tables~\ref{tab:ephem1} and \ref{tab:ephem2}. The residuals used to create the plot are the same as in Figure~\ref{fig:timing_res}.}
\label{fig:timing_res_orb}
\end{figure*}

Following the procedures detailed in Section~\ref{ssec:data_proc}, we obtained the first phase-coherent timing solutions for M2A--G. Additionally, we performed Bayesian timing analysis to refine the measurements of timing model parameters, using the \texttt{TEMPONEST} software package \citep{Len+14}. On top of the timing parameters listed in Table~\ref{tab:ephem1} and \ref{tab:ephem2}, we included two parameters to model white noise in the data: a multiplicative factor $E_{\rm f}$ (\texttt{EFAC}) to account for possible underestimation of TOA uncertainties, and a factor $E_{\rm q} $(\texttt{EQUAD}) added in quadrature to incorporate other additive noise such as pulse phase jitter and systematics in the data \citep[e.g.][]{lvk+11,lkl+12}. These two parameters are related to the uncertainty of the TOA measurement, $\sigma_{\rm r}$, as follows:
\begin{equation}
    \sigma=\sqrt{E^2_{\rm q}+E^2_{\rm f}\sigma^2_{\rm r}}.
    \label{eq:whitenoise}
\end{equation}
We also searched for low-frequency noise but did not detect its presence in the dataset. The measurements of the timing-model parameters, as well as the derived parameters, are given in Table~\ref{tab:ephem1} and \ref{tab:ephem2}. Figure~\ref{fig:timing_res} displays the timing residuals of the M2 pulsars after subtracting the best-estimated timing models, and the same residuals plotted as a function of the orbital phase are shown in Figure~\ref{fig:timing_res_orb}. The achieved rms of the residuals falls in the range of 1--100\,$\mu$s, with the lowest value of 3.5\,$\mu$s from M2C. 

\begin{figure*}
\hspace*{-1.1cm}
\includegraphics[scale=0.6]{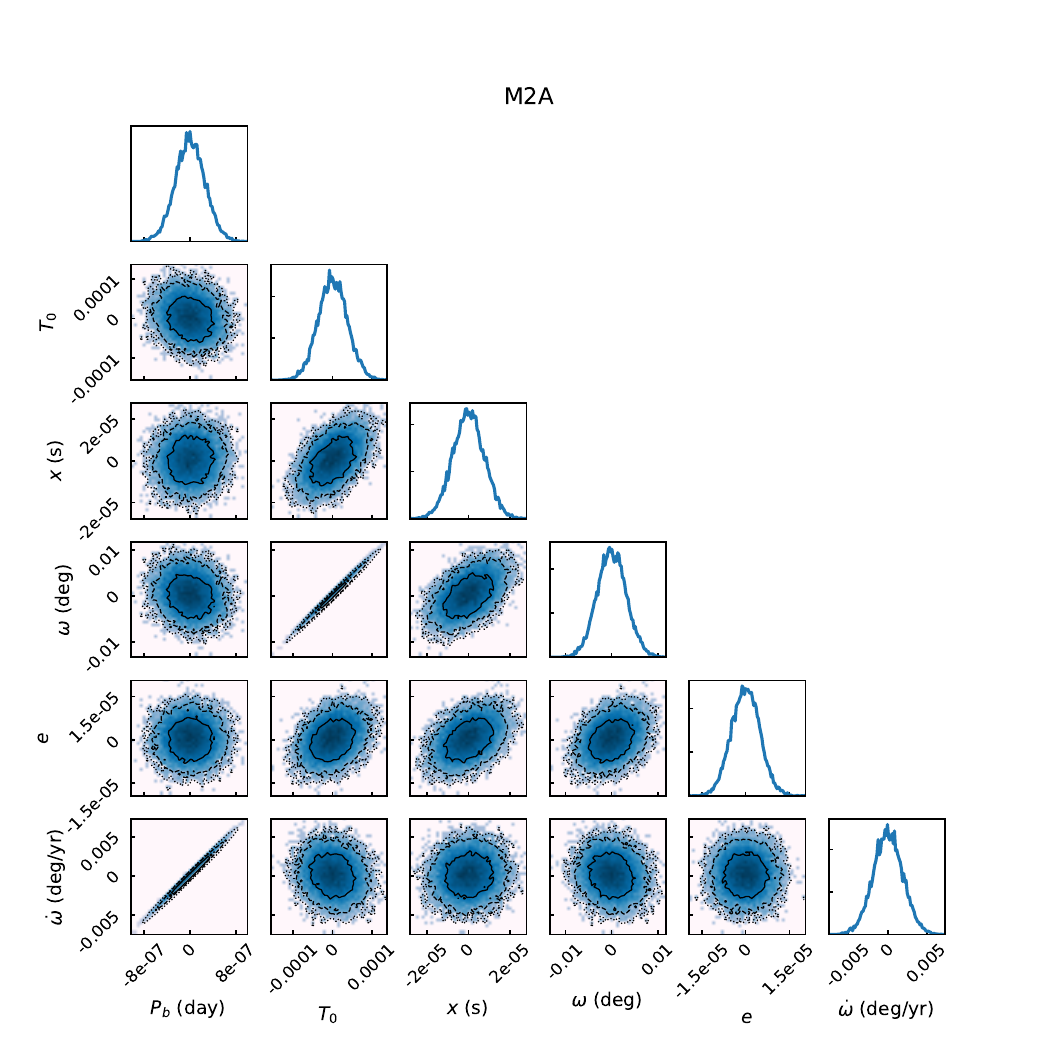}
\hspace*{-1.1cm}
\includegraphics[scale=0.6]{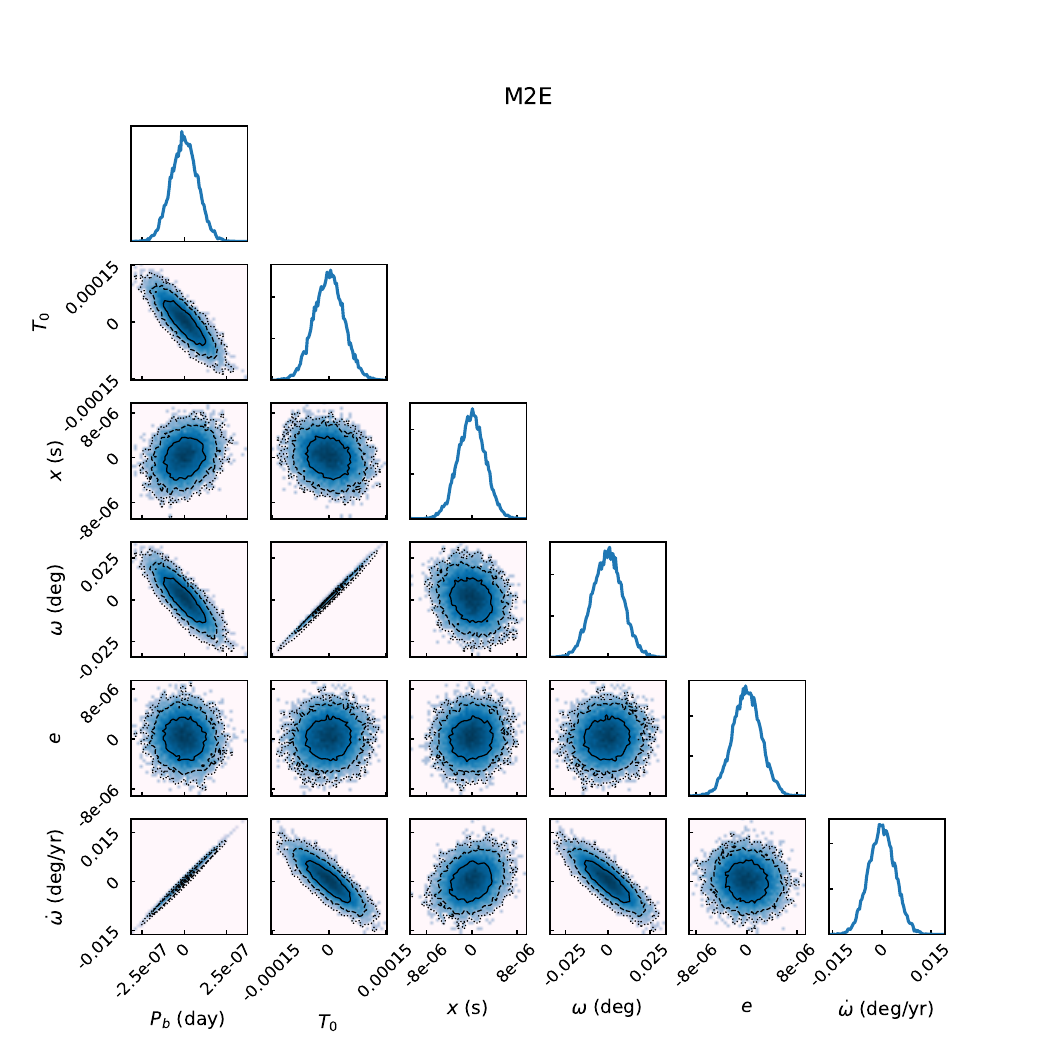}
\caption{Posterior distributions of binary parameters from the Bayesian timing analysis of M2A and M2E. The values shown here were after subtracting the median values in Table~\ref{tab:ephem1} and \ref{tab:ephem2}.}
\label{fig:pd}
\end{figure*}

We have obtained significant measurements of the spin frequency derivative for all M2 pulsars. For M2B and M2C, the spin frequency derivative is positive, which is likely attributed to the cluster potential rather than being intrinsic to the pulsars themselves (see Section~\ref{ssec:cluster_potential} for more detailed investigations). In parallel, we have obtained significant or tentative measurements of proper motion in RA for M2B, M2C, M2E, and in DEC for M2D. For the rest of the pulsars, the proper motion terms were not well constrained with our current dataset. Nonetheless, the measured proper motion in RA for M2C is highly consistent with that reported for the M2 cluster, 3.445~$\pm$~0.009~mas~yr$^{-1}$ \citep{vb21}. The constraints obtained in other pulsars also show a broad consistency. The Keplerian orbital parameters of the M2 pulsars are generally well measured. We have detected significant orbital eccentricity for M2A, M2B, M2D, and M2E. For the orbit of M2A and M2E, we measured the advance of periastron to be 0.026(2) and 0.135(4)\,deg/yr, respectively, corresponding to total system masses of the systems to be 1.75(13) and 1.80(5)\,M$_{\rm \odot}$, respectively. The posterior distributions of the orbital parameters from the Bayesian timing analysis of M2A and M2E can be found in Figure~\ref{fig:pd}. 

\begin{table*}[!htbp]
\centering
\caption{Measured and derived timing-model parameters of the M2 pulsars. Here the values are the medians of the posterior distributions from the Bayesian timing analysis. The brackets represent the 1-$\sigma$ credible interval of the respective parameters. The proper motion terms are defined as: $\mu_\alpha = \dot{\alpha}\cos\delta$, $\mu_\delta=\dot{\delta}$. Note that the reduced $\chi^2$ of the fit are all close to unity, after inclusion of the noise modeling. }
\label{tab:ephem1}
\hspace*{-1.8cm}
\begin{tabular}[c]{lllll}
\hline
Pulsar & M2A & M2B & M2C & M2D \\
\hline
MJD range & 58795--60512 & 58795--60512 & 58795--60512 & 58795--60512 \\
Number of TOAs & 244 & 241 & 246 & 247 \\
rms timing residuals ($\mu$s) & 57.5 & 21.3 & 3.52 & 19.0 \\
Reference epoch (MJD) & 59514.5 & 59498.5 & 58805.4 & 58852.3 \\
\hline
\multicolumn{5}{c}{Measured parameters} \\
\hline
Right Ascension, $\alpha$ (J2000) & 21:33:27.0743(3) & 21:33:26.7651(7) & 21:33:28.7789(1) & 21:33:27.1756(6) \\
Declination, $\delta$ (J2000) & $-$00:49:12.30(1) & $-$00:48:50.46(3) & $-$00:49:36.546(4) & $-$00:49:33.12(2)\\
Spin frequency, $\nu$ (Hz) & 98.52910098448(3) & 143.37847196762(2) & 332.78628361061(4) & 237.20655892966(8) \\
Spin frequency derivative, $\dot{\nu}$ (s$^{-2}$) & $-$1.6340(4)$\times10^{-15}$ & 1.0509(8)$\times10^{-15}$ & 2.4532(5)$\times10^{-15}$ & $-$4.2864(10)$\times10^{-15}$ \\
DM (cm$^{-3}$~pc) & 43.444(6) & 43.726(2) & 44.074(3) & 43.570(2) \\
DM1 (cm$^{-3}$~pc~yr$^{-1}$) & --- & --- & -0.005(1) & --- \\
DM2 (cm$^{-3}$~pc~yr$^{-2}$) & --- & --- & 0.0010(4) & --- \\
Proper motion in $\alpha$, $\mu_{\rm\alpha}$ (mas\,yr$^{-1}$) & --- & 7(3) & 3.4(4) & 3(3) \\
Proper motion in $\delta$, $\mu_{\rm\delta}$ (mas\,yr$^{-1}$) & --- & 6(8) & -1(1) & 13(5) \\
Binary model & DD & ELL1 & ELL1 & ELL1 \\
Orbital period, $P_{\rm b}$ (d) & 4.2554873(3) & 9.34712630(2) & 1.1091193455(3) & 3.429704477(3) \\
Projected semi-major axis, $x$ (s) & 3.140014(8) & 5.891176(3) & 1.0966404(6) & 3.720671(2) \\
Longitude of periastron, $\omega$ (deg) & 319.079(3) & --- & --- & ---\\
Epoch of periastron (MJD), $T_{\rm 0}$ & 59482.95666(4) & --- & --- & ---\\
Orbital eccentricity, $e$ & 0.075158(5) & ---  & --- & --- \\
Epoch of ascending node (MJD), $T_{\rm asc}$ & --- & 58856.491016(2) & 58799.6909440(2) & 58799.0569822(8) \\
$\hat{x}$ component of the eccentricity, $\kappa$ & --- & $-$2.25(9)$\times10^{-5}$ & 9(8)$\times10^{-7}$ & $-$4(1)$\times10^{-6}$ \\
$\hat{y}$ component of the eccentricity, $\eta$ & --- & 6.7(9)$\times10^{-6}$ & 10(8)$\times10^{-7}$ & 6(1)$\times10^{-6}$ \\
Advance of periastron, $\dot{\omega}$ (deg/yr) & 0.026(2) & ---  & --- & --- \\
\hline
\multicolumn{5}{c}{Derived parameters} \\
\hline
Spin period, $P$ (s) & 0.010149285744092(3) & 0.006974547756554(1) & 0.0030049315408987(3) & 0.004215735030736(1)\\
Spin period derivative, ($\dot{P}$) & $1.6832(4)\times10^{-19}$ & $-5.112(4)\times10^{-20}$ & $-2.2152(4)\times10^{-20}$ & $7.618(2)\times10^{-20}$ \\
Mass function, $f(M_{\rm p})$ (${ M}_\odot$) & 0.00183560(1) & 0.002512653(3) & 0.001151115(2) & 0.004701466(9) \\
Orbital eccentricity, $e$ & --- & $2.34(9)\times10^{-7}$ & $1.3(8)\times10^{-6}$ & $8(1)\times10^{-6}$ \\
Total mass, $M_{\rm tot}$ ($M_{\odot}$) & 1.75(13) & --- & --- & --- \\
Min. companion mass, $M_{\rm c, min}$ ($M_{\odot}$) &0.1612 & 0.1806 & 0.1365 & 0.227 \\
Median companion mass, $M_{\rm c, med}$ ($M_{\odot}$) & 0.1884 & 0.2113 & 0.1592 & 0.2664 \\
Max. companion mass, $M_{\rm c, max}$ ($M_{\odot}$) &0.4094 & 0.4639 & 0.3413 & 0.5998 \\
\hline
\end{tabular}
\end{table*}

\begin{table*}[!htbp]
\centering
\caption{Continuation of Table~\ref{tab:ephem1}.}
\label{tab:ephem2}
\hspace*{-0.47cm}
\begin{tabular}[c]{llll}
\hline
Pulsar & M2E & M2F & M2G \\
\hline
MJD range & 58795--60512 & 58795--60512 & 58795--60512\\
Number of TOAs & 216 & 82 & 82 \\
rms timing residuals ($\mu$s) & 20.1 & 64.6 & 11.7 \\
Reference epoch (MJD) & 59486.5 & 59490.5 & 59763.9 \\
\hline
\multicolumn{4}{c}{Measured parameters} \\
\hline
Right Ascension, $\alpha$ (J2000) & 21:33:26.9486(6) & 21:33:27.9850(6) & 21:33:26.7284(2)  \\
Declination, $\delta$ (J2000) & $-$00:49:28.53(2) & $-$00:49:10.79(3) & $-$00:49:48.399(6)        \\
Spin frequency, $\nu$ (Hz) & 270.04262113057(8) & 209.16616986511(7)  & 394.36226958417(3)       \\
Spin frequency derivative, $\dot{\nu}$ (Hz$^{-1}$) & $-$2.861(1)$\times10^{-15}$ & $-$1.331(2)$\times10^{-15}$ & $-$4.1735(10)$\times10^{-15}$ \\
DM (cm$^{-3}$~pc) & 43.651(2)  & 43.33(1) & 43.413(2) \\
Proper motion in $\alpha$, $\mu_{\rm\alpha}$ (mas\,yr$^{-1}$) & 5(2)  & ---  & ---   \\
Proper motion in $\delta$, $\mu_{\rm\delta}$ (mas\,yr$^{-1}$) & 3(6)  & ---  & ---    \\
Binary model & DD & ELL1 & ELL1 \\
Orbital period, $P_{\rm b}$ (d) & 1.59730248(8) & 3.59846939(5) & 0.120357703(1)    \\
Projected semi-major axis, $x$ (s) & 1.841753(2) & 1.61880(1) & 0.025324(2)    \\
Longitude of periastron, $\omega$ (deg) & 197.584(9) & --- & --- \\
Epoch of periastron (MJD), $T_{\rm 0}$ & 58811.53308(4) & --- & --- \\
Orbital eccentricity, $e$ & 0.035730(2) & --- & --- \\
Epoch of ascending node (MJD), $T_{\rm asc}$ & --- & 58803.09859(1) & 59763.902858(2)    \\
$\hat{x}$ component of the eccentricity, $\kappa$ & --- & $-$5.5(2)$\times10^{-6}$ & 5.5(3)$\times10^{-5}$   \\
$\hat{y}$ component of the eccentricity, $\eta$ & ---   & 2(2)$\times10^{-5}$ & $-$0.0004(2)  \\
Advance of periastron, $\dot{\omega}$ (deg/yr) & 0.135(4) & --- & --- \\
\hline
\multicolumn{4}{c}{Derived parameters} \\
\hline
Spin period, $P$ (s) & 0.003703119143983(1) & 0.004780887849335(15) & 0.0025357395398258(2) \\
Spin period derivative, ($\dot{P}$) & $3.923(2)\times10^{-20}$ & $3.042(5)\times10^{-20}$ & $2.6836(6)\times10^{-20}$ \\
Mass function ($M_{\odot}$) & 0.00262907(1) & 0.000351743(9) & 0.0000012039(3) \\
Orbital eccentricity, $e$ & --- & $2(2)\times10^{-5}$ & 0.0004(2) \\
Total mass, $M_{\rm tot}$ ($M_{\odot}$) & 1.80(5) & --- & --- \\
Min. companion mass, $M_{\rm c, min}$ ($M_{\odot}$) & 0.1835 & 0.09002 & 0.01308 \\
Median companion mass, $M_{\rm c, med}$ ($M_{\odot}$) & 0.2148 & 0.1046 & 0.0151 \\
Max. companion mass, $M_{\rm c, max}$ ($M_{\odot}$) & 0.4724 & 0.2186 & 0.0303 \\
\hline
\end{tabular}
\end{table*}

\subsection{Scintillation study of M2C} \label{ssec:scint}

\begin{table*}[!ht]
\caption{The scintillation parameters of M2C at 1250\,MHz. Here $\Delta\tau_{\rm d}$ and $\Delta\nu_{\rm d}$ are the scintillation timescale and bandwidth, respectively, $\sigma_{r}$ is their fractional error, $u$ is the scintillation strength, $\eta$ is the arc curvature, and $D_{\rm kpc}$ is the pulsar distance.}\label{tab:scintillation parameters}
\hspace*{-2cm}
\centering
\begin{tabular}{cccccccc}
\hline \hline
MJD & $\Delta\tau_{\rm d}$ & $\Delta\nu_{\rm d}$ & $\sigma_{r}$ & $u$  & $\eta$ & $s$ & $D_{\rm kpc}(1-s)$ \\ 
& (min) & (MHz)&  & &(${\rm m}^{-1}\,{\rm mHz}^{-2}$) & & {\rm kpc} \\ 
\hline
59555  &12.58$\pm$0.21  &3.77$\pm$0.24  & 0.07 & 26  & 4540$\pm$234 & 0.53$\pm$0.03 & 5.50$\pm$0.35 \\ 
59662 &11.35$\pm$0.18  &5.79$\pm$0.46  & 0.08 & 21  & 5412$\pm$426 & 0.57$\pm$0.04 & 5.03$\pm$0.47 \\ 
59679 &12.77$\pm$0.24  &3.34$\pm$0.31  & 0.06 & 27  & 3440$\pm$106 & 0.48$\pm$0.01 & 6.08$\pm$0.12  \\
59764 &11.63$\pm$0.31  &3.47$\pm$0.24  & 0.06 & 27  & 2842$\pm$151 & 0.48$\pm$0.01 & 6.08$\pm$0.12 \\
\hline
\end{tabular}
\end{table*}

\begin{figure*}
\hspace*{-1.0cm}
\includegraphics[scale=0.42]{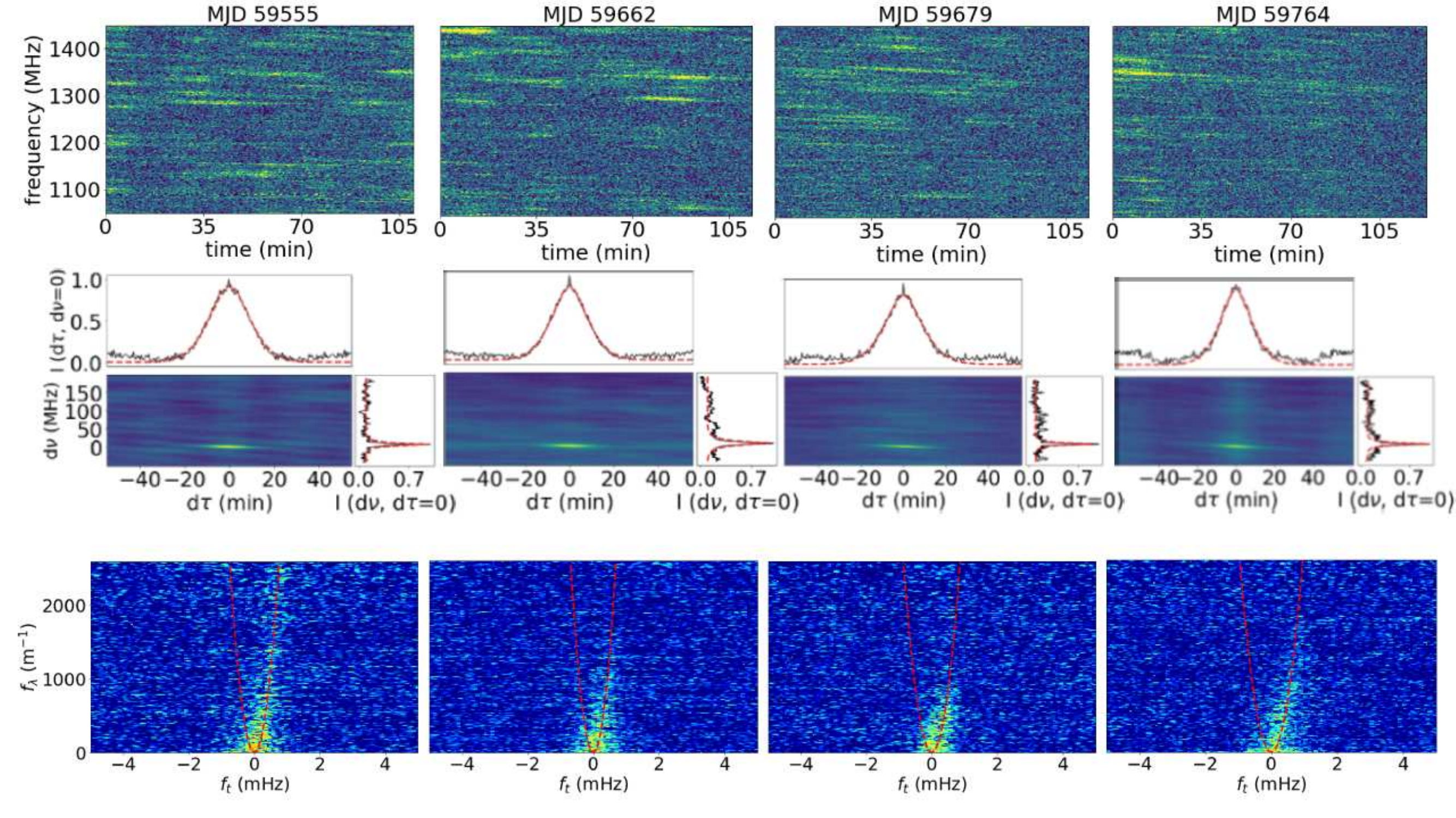}
\caption{Dynamic spectra (top panels), ACFs (middle panels), and secondary spectra (bottom panels) of M2C in MJDs 59555, 59662, 59679, and 59764, respectively. The data lengths are 109, 113, 113, and 123\,min, respectively, while the bandwidth is 400\,MHz. The sub-integration time and frequency resolution are 30\,s and 0.98\,MHz, respectively. In the side panels of the ACF, the red lines show the best-fit results for either $\Delta \tau_d$ or $\Delta \nu_d$. In secondary spectra, the red dotted line represents the best-fit arc curvature. \label{fig:scintillation}}
\end{figure*}

As shown in Figure~\ref{fig:Flux}, the flux density of the M2 pulsars varies significantly on an epoch-by-epoch basis. The intensity and time scale of this variation are typical as expected from diffractive interstellar scintillation \citep{ric90}. Investigating this phenomenon serves as a valuable tool for understanding the ISM along the LoS to the cluster \citep[e.g.,][]{lzy+24}. By examining the dynamic spectrum, which is a map of pulsar intensity as a function of observing frequency $\nu$ and time $t$, scintillation parameters, such as its time and frequency scale, can be derived to investigate the turbulence process within the ISM \citep[e.g.,][]{wmj+05}. The secondary spectrum, defined as the squared modulus of the two-dimensional Fourier transform of the dynamic spectrum, often displays an arc structure. This arc structure can be utilized to model the geometry and structure (e.g., scale, distance) of the scattering screen along the LoS to the cluster \citep[e.g.,][]{yzw+20}.

Achieving a detailed study of the dynamic spectrum typically requires high S/N of detection from individual epoch observations, which is challenging given the low flux densities of the M2 pulsars. Through visual inspection of the dynamic spectra of the M2 pulsars, we managed to detect clear scintillation features in four epoch observations (MJD 59555, 59662, 59679, and 59764) of M2C, as shown in Figure~\ref{fig:scintillation}. Based on these dynamic spectra, we calculated the two-dimensional auto-correlation functions (ACF), using the \textsc{scintools} software package \citep{rea20}. The ACF is defined as: 
\begin{equation}
C(\Delta\nu,\Delta\tau) = \sum_{\nu}{}\sum_{t}{} \Delta D(\nu, t) \Delta D(\nu + \Delta\nu, t + \Delta\tau),
\end{equation}
where $D(\nu, t)$ is the pulsar intensity as a function of frequency, $\nu$ and time, $t$. $\Delta D(\nu, t)$ = $D(\nu, t) - \overline{D}$ (where ``${-}$" denotes averaging). The normalized ACF is $N(\Delta\nu,\Delta\tau) = C(\Delta\nu,\Delta\tau)/C(0,0)$. Following \citet{rch+19}, the scintillation timescale $\Delta\tau_{\rm d}$ and the scintillation bandwidth $\Delta\nu_{\rm d}$ were derived from
\begin{eqnarray}
{\rm C}(\Delta\tau,0) &=&{\rm A\ exp} (- |\frac{\Delta\tau}{\Delta\tau_{\rm d}}|^{\frac{5}{3}})\label{eq:2}, \\
{\rm C}(0,0) &=& {\rm A + W}, \\
{\rm C}(0,\Delta\nu) &=& {\rm A\ exp} (- |\frac{\Delta\nu}{\Delta\nu_{\rm d}/{\rm ln2}}|),
\end{eqnarray}
where $A$ is a constant to be fitted and $W$ is the noise spike. The fractional error of the scintillation parameters is defined as \citep{brg99},
\begin{eqnarray}
\sigma_r = \left(\frac{\mathit{f}BT}{\Delta\nu_{\rm d}\Delta\tau_{\rm d}}\right)^{-0.5},
\end{eqnarray} 
where $T$ and $B$ are the data length and bandwidth, respectively. The filling factor $\mathit{f}$ is the percentage of scintillation area in the dynamic spectrum. As in \citet{lzy+24}, we calculated the percentage of data with more than 1-$\sigma$ in the dynamic spectra and chose $\mathit{f}$ = 0.25. The scintillation strength parameter is defined as \citep{brg99} 
\begin{eqnarray}
u \approx \sqrt{\frac{2\nu}{\Delta\nu_{\rm d}}},
\end{eqnarray} 
where $\nu$ is the central frequency of the data, and $u>1$ is defined as strong scintillation, while $u<1$ indicates weak scintillation. The measurements of scintillation bandwidth and timescale are summarized in Table~\ref{tab:scintillation parameters}. It can be seen that the scintillation timescale and bandwidth of M2C are generally consistent across different epochs, with values of approximately 12\,min and 4\,MHz, respectively. The scintillation strength from these observations is well above unity (ranging from 21 to 27), indicating strong scintillation along the LoS of M2C.

The secondary spectrum is calculated by performing a two dimensional Fourier transform of the dynamic spectrum \citep[e.g.,][]{rea20}:
\begin{eqnarray}
P(f_{t},f_{\lambda}) =10\log_{10}(\mid\tilde{D}(t,\lambda)\mid^{2}),
\end{eqnarray}
where $f_{t}$ and $f_{\lambda}$ are the conjugate time and the conjugate wavelength, respectively. The arc structure in the secondary spectrum is described by \citep[e.g.,][]{rch+19}:
\begin{eqnarray}
f_{\lambda}=\eta f_{t}^{2},
\end{eqnarray}
in which
\begin{eqnarray}
\label{con:inventoryflow}
\eta   =  1.543 \times 10^{7} \frac{D_{\rm kpc}s(1-s)}{(V_{{\rm eff},\perp}\cos\psi)^{2}}. \label{eq:9}
\end{eqnarray}
Here, $\eta$ is the so-called arc curvature, $D_{\rm kpc}$ is pulsar distance in kpc, $s$ is the relative distance of the scattering screen to the pulsar, and $\psi$ is the angle between the major axis of the anisotropic structure and the effective velocity vector. The effective perpendicular velocity in~${\rm km\ s^{-1}}$, $V_{{\rm eff},\perp}$, can be written as 
\begin{equation}
    V_{{\rm eff},\perp} = (1-s)V_{{\rm psr},\perp}+sV_{{\rm E},\perp}-V_{{\rm scr},\perp},
\end{equation}
where $V_{{\rm psr},\perp}$, $V_{{\rm E},\perp}$ and $V_{{\rm scr},\perp}$ are the transverse velocity of the pulsar, Earth and scattering medium with respect to the solar-system barycenter (SSB), respectively. As shown in Figure~\ref{fig:scintillation}, we detected clear scintillation arcs in the secondary spectra from the four epoch observations of M2C. With \textsc{scintools}, the arc curvatures were measured to be 4540$\pm$234, 5412$\pm$426, 3440$\pm$106, and 2842$\pm$151\,m${^{-1}}$\,mHz$^{-2}$, respectively. These measurements suggest some variation in the scintillation characteristics. It is important to note that M2C is in a binary system, and the variation in the scintillation arc may be influenced by both the annual and the orbital motion of the pulsar \citep{rcb+20}. However, the number of detections is insufficient to model either the annual or the orbital effects causing the variation in the scintillation arc.

Given that the measured proper motion of M2C (see Table~\ref{tab:ephem1}) is consistent with that of the cluster (3.445$\pm$0.009~mas/yr for RA and $-2.177 \pm0.009$~mas/yr for DEC), we can use the proper motion and distance (11.69\,kpc) of M2 \citep{vb21} to approximately evaluate the pulsar's transverse velocity, $V_{\rm psr}$. This resulted in $V_{\rm psr}$ being 191(1)\,${\rm km\ s^{-1}}$ in RA and $-121(1)\,{\rm km\ s^{-1}}$ in DEC. Meanwhile, the Earth's transverse velocity $V_{\rm E}$, was calculated to be $-$11.96, $-$21.34, $-$14.96, and 21.42\,${\rm km\ s^{-1}}$ for RA, and 2.59, $-$11.63, $-$10.86, and 2.60\,${\rm km\ s^{-1}}$ for DEC, for MJD 59555, 59662, 59679, and 59764, respectively. Using Equation~\ref{eq:9} and assuming $V_{{\rm scr},\perp}=0$, $\psi=0$, we calculated the relative distance of the scattering screen to the pulsar $s$, using the curvature of the scintillation arc measured in the secondary spectrum in each epoch. The values are reported in Table~\ref{tab:scintillation parameters}, along with the approximate position of the scattering screen, $D_{\rm kpc}(1-s)$. It can be seen that the scattering screen is located approximately halfway between M2C and the Earth, at a distance of about 5--6\,kpc.


\section{Discussions} \label{sec:dis}
\subsection{Properties of pulsars in cluster M2}\label{sec:Pro_M2}
So far, all seven pulsars in M2 are binary millisecond pulsars, which is consistent with the expectation from a non-core-collapsed cluster. Typically, the core density level of a cluster is evaluated using the King-model central concentration parameter, defined as $c=\log(r_t/r_c)$, where $r_t$ is the tidal radius and $r_c$ is the core radius of the cluster. For M2, $c\approx1.59$, which is a typical value for a non-core-collapsed cluster according to the King model \citep{har10a}. Meanwhile, a more direct evaluation of the fraction of binaries in a cluster is the single-binary encounter rate, which will be discussed in Section~\ref{ssec:encounter rate}.

The pulsars M2A to M2F have orbital periods ranging from 1 to 10\,day and median companion mass between 0.1 and 0.2\,M$_{\odot}$. According to classical binary evolution theories \citep[e.g.,][]{tv23}, these are likely to be neutron stars --- white dwarf binaries that evolved from low-mass X-ray binary systems. Meanwhile, M2G has a tight orbit of 2.9\,hr and a maximum companion mass (with 90\% probability) of 0.03\,\,M$_{\odot}$, which are typical characteristics of a black widow system. We examined the time variability of flux density from a few epoch observations with the highest S/N of detection for M2G and did not find clear evidence for eclipses over the course of an orbit. This could be due to the low S/N in each sub-integration, which prevents us from clearly distinguishing between the on and off status. Meanwhile, from Figure~\ref{fig:timing_res_orb} there is also no clear offset in the timing data around the super conjunction (at orbital phase of 0.25) of the pulsar orbit. Thus, it is more likely that M2G is a non-radio-eclipsing black widow pulsar, similar to several other systems identified to date \citep[e.g.,][]{bjs+20,bgv+25}. One common feature of these systems is that, like M5G, (the orbital period is about 0.11 days, and the minimum companion mass is about 0.02 $M_\odot$ \citep{Zha+23}), their mass functions are relatively small for black widow systems, which suggests that they are being observed at lower orbital inclinations than the eclipsing black widow systems \citep{bgv+25}. However, such systems seem to have a smaller variability in their orbital periods, suggesting that there is a physical difference, not merely associated with the viewing angle. For M2G, this hypothesis may be tested later through long-term timing analysis to check for any orbital variations in the binary system.

\subsection{Search for multi-wavelength counterparts of M2 pulsars} 

\begin{figure*}
    \centering
	\includegraphics[width=1.0\linewidth]{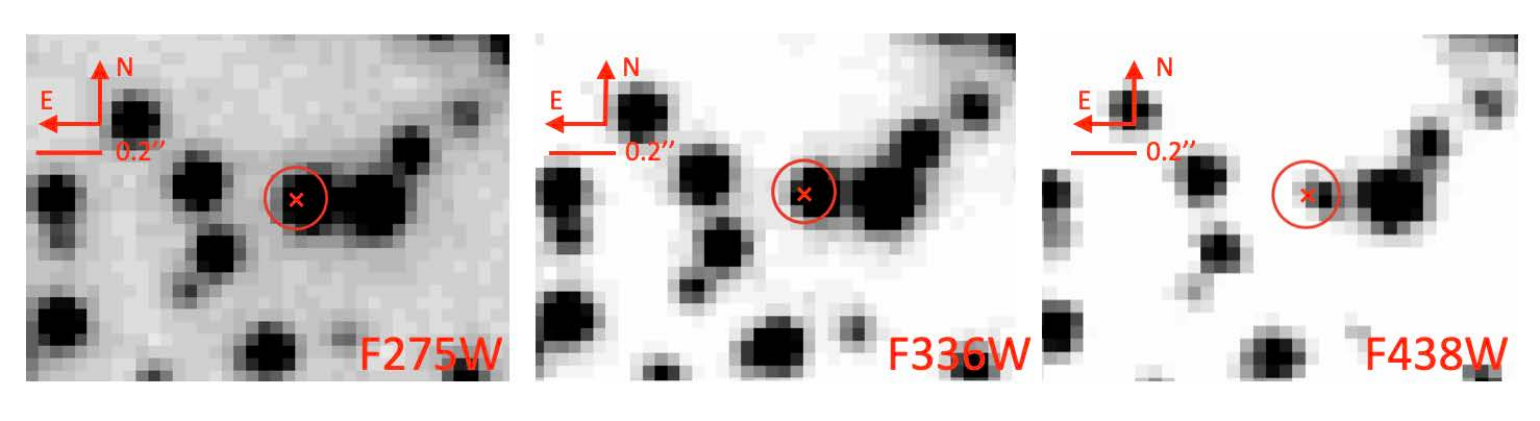}
	\caption{Finding chart of the regions surrounding the position of M2C in F275W, F336W, and F438W images, respectively. The red crosses represent the radio position and the red circles have a radius of 0.1\,arcsec which is approximately three times the maximum position error of the M2 pulsars shown in Table~\ref{tab:ephem1} and \ref{tab:ephem2}. The star within the red circle is the identified companion of M2C. \label{fig:hst_m2c}}
\end{figure*}

From the mass functions of the binary systems, the companions of the known M2 pulsars are expected to be white dwarfs or low-mass stars, which could be detected in the optical band. Accordingly, we searched for optical counterparts of the known M2 pulsars using data from the Hubble Space Telescope UV Globular Cluster Survey (HUGS, GO-10775, PI: Sarajedini, and GO-13297, PI: Piotto)\footnote{\url{https://archive.stsci.edu/prepds/hugs/\#dataaccess}}. The dataset contains a selection of multi-wavelength imaging observations spanning the near-ultraviolet to optical regimes, specifically including the following Hubble Space Telescope (HST) filters: F275W, F336W, F438W, F606W, and F814W. More detailed information on the dataset and its calibration process can be found in \citet{pmb+15} and \citet{nlp+18}. We cross-matched potential counterparts in HST optical images using the pulsar positions provided in Tables~\ref{tab:ephem1} and \ref{tab:ephem2}. The resolution of the HST image is approximately 0.02\,arcsec, while the 1-$\sigma$ credible intervals of the pulsar positions are approximately 0.03\,arcsec or below. 

\begin{figure}
    \centering
	\includegraphics[width=\columnwidth]{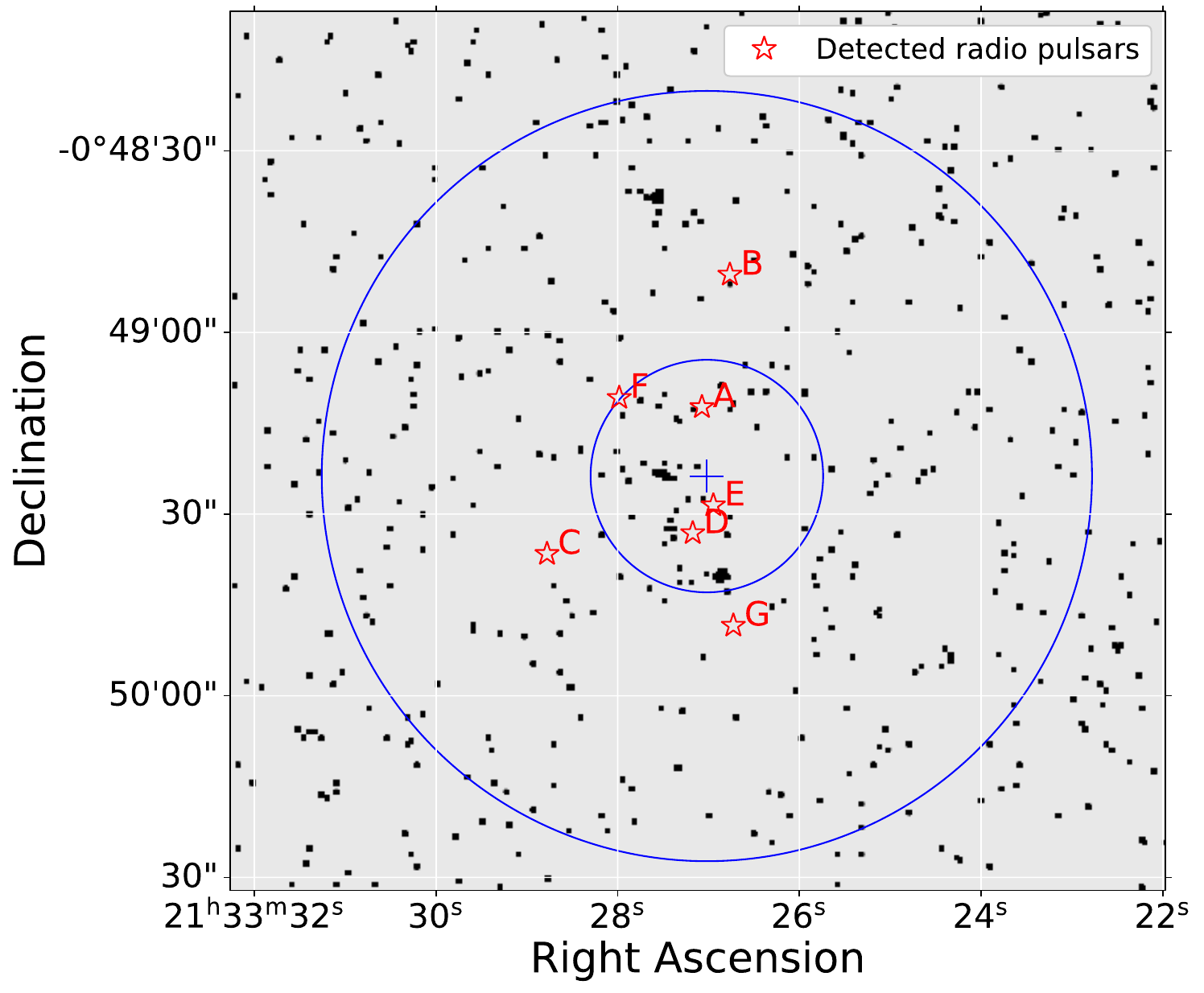}
	\caption{X-ray image of the M2 region from the Chandra archival data, where the positions of the known M2 pulsars are shown as red stars. The center of M2 is marked by a blue cross. The small blue circle represents the core radius of the cluster, $0.32'$. The larger blue circle shows the half-light radius of the cluster, $1.06'$.}
	\label{fig:X_ray_sources}
\end{figure}

To ensure robust counterpart identification, we adopted a matching radius of either 0.02\,arcsec (equivalent to the HST point-spread function) or 3 times the pulsar's positional uncertainty (3-$\sigma$), whichever was larger, centered on each pulsar's timing-derived coordinates. With this criterion, we detected an apparent optical counterpart of M2C, located at a separation of 0.008\,arcsec from its radio timing position. The source exhibits magnitudes of 19.98(6), 20.374(4), and 21.4(1) in F275W, F336W and F438W filters, respectively. The finding charts in three different wavelengths are shown in Figure~\ref{fig:X_ray_sources}. The color-magnitude diagrams (CMDs) are presented in Figure~\ref{fig:m2c_cooling}. This companion star is located on the blue side of the white-dwarf cooling sequence, indicating that it is a hot white dwarf.



Typically, rotation-powered MSPs generate faint ($\le~10^{33}$~erg~s$^{-1}$) X-ray emission with a soft X-ray spectrum \citep{bgh+06}. The X-ray emission can originate both thermally from the surface of the neutron star and non-thermally from the pulsar's magnetosphere. Both thermal and non-thermal emissions can be modulated by the pulsar's rotation, leading to pulsations in the X-ray light curves. Non-thermal X-ray radiation can also be modulated by the binary orbital period, a phenomenon observed thus far only in the so-called ``spider'' pulsar systems, such as Terzan 5 A, P, and O \citep{bbh+21}. This modulation can be explained as non-thermal radiation caused by the interaction between the pulsar wind and the materials of the companion \citep{at93}. In the case of M2, M2G is expected to be a black widow pulsar, a subclass of spider pulsar system, see Section~\ref{sec:Pro_M2}. 

Using the timing positions of M2A to M2G from Tables~\ref{tab:ephem1} and \ref{tab:ephem2}, we searched the Chandra archival data \footnote{{\url{https://cxcfps.cfa.harvard.edu/cda/footprint}}} to locate the X-ray counterparts of these pulsars. The energy range of the background X-ray image is from 0.2 to 20\,keV. The X-ray image of the M2 region is shown in Figure~\ref{fig:X_ray_sources}, with the positions of the M2 pulsars marked. The error in the timing positions is significantly smaller than the resolution of the image. Within a 3-$\sigma$ range of the pulsar positions, we did not find clear evidence for the presence of X-ray counterparts for the M2 pulsars.

\subsection{Estimations from binary evolution} 

\begin{figure}
    \centering
	\includegraphics[width=\columnwidth]{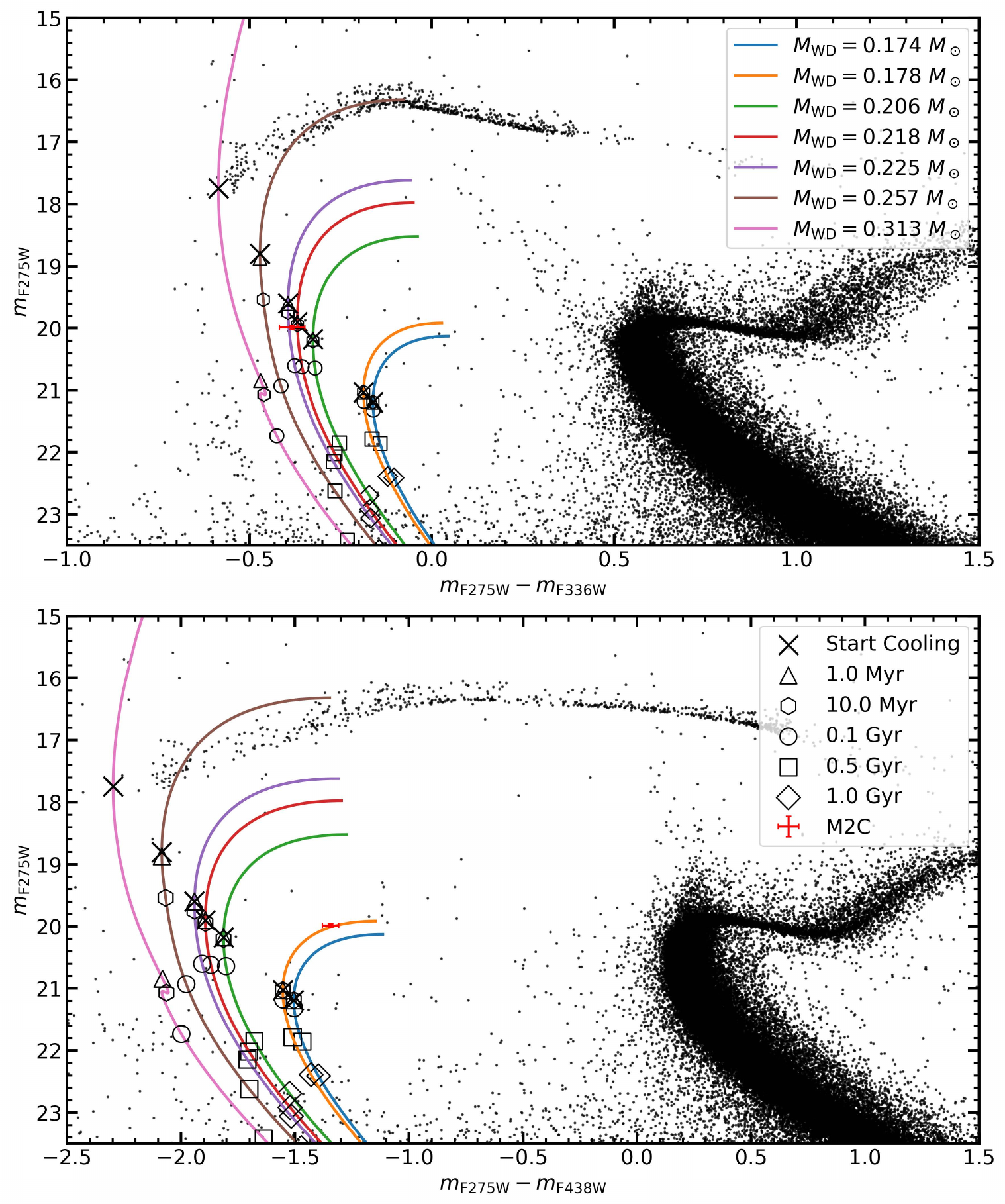}
	\caption{Top panel: CMD of M2C in a combination of the F275W and F336W filters. Bottom panel: CMD of M2C in a combination of the F275W and F438W filters. In both panels, the position of the companion star to M2C is marked with a red cross and the error bars correspond to the 1$\sigma$ confidence level uncertainties. The colored curves are helium white-dwarf tracks with a selection of masses. The different cooling ages are marked by a variety of symbols.}
	\label{fig:m2c_cooling}
\end{figure}

To characterize the white-dwarf companion of M2C, we calculated the binary evolution track of the system and compared the results with our observations. The metallicity and distance of M2 were used as $[{\rm Fe/H}] = -1.65$ and $11.69~\rm kpc$, respectively \citep{blt+11,lbv+22}. We used the \textsc{Astrolib PySynphot} software package\footnote{\url{https://pysynphot.readthedocs.io/en/latest/index.html}} to convert theoretical luminosities into absolute HST magnitudes. Adopting a distance modulus $(m-M)_0 = 15.26$, a color excess $E(B-V) = 0.06$, and extinction laws from \citet{ccm89}, we calibrated the theoretical absolute magnitudes to the observational frame of M2 \citep{hlm+18,sjw+25}. Binary evolution models were computed using \textsc{MESA}\footnote{\url{https://docs.mesastar.org/en/latest/}  The MESA inlists and input model of this work are available on Zenodo: \url{https://doi.org/10.5281/zenodo.15778913}.  } \citep[version 24.08.1,][]{pbd+11,pca+13,pms+15,psb+18,pss+19}, initialized with a $1.35~M_\odot$ neutron star and a $1.0~M_\odot$ companion at metallicity $Z = 0.0003$. The resulting theoretical cooling tracks for helium white dwarfs of various masses are presented in Figure~\ref{fig:m2c_cooling}.

The masses of the helium white dwarfs are estimated to be $\sim0.225~M_\odot$ or $\sim0.178~M_\odot$ based on the best-fit model tracks shown in Figure~\ref{fig:m2c_cooling}. These align are consistent with the mass range derived from radio timing observations.
The formation of these helium white dwarfs by single-star evolution is almost impossible.
The mass function and the low eccentricity suggest that this object is the companion of M2C.
Taking the two mass estimates of the white dwarf and assuming an orbital inclination of 60\,deg, the corresponding pulsar mass is 2.3 and 1.6\,$M_{\odot}$, respectively. The cooling age of M2C's companion presents another notable result. As shown in the top panel of Figure~\ref{fig:m2c_cooling}, the estimated cooling age is approximately $\sim10$ Myr. In the case of the bottom panel, the companion has even not started cooling yet. In either interpretation, the companion of M2C emerges as an exceptionally young helium white dwarf, with an age substantially younger than that of M2 itself, which is $\sim 12.5$ Gyr \citep{mmp+15}. It is also approximately an order of magnitude younger than the white-dwarf companions of PSR~B1310+18A in the GC M53 \citep{lfc+25} and PSR~J1835$-$3259B in the bulge GC NGC 6652 \citep{grf+22,ccp+23}; this makes the companion of M2C the youngest white dwarf in a binary system among all GCs. Interestingly, this system has also an extraordinarily small eccentricity for an MSP and a helium white-dwarf system in a GC (see Table~\ref{tab:ephem1}), which agrees with the idea that the recycling ended very recently and there has been no time for interactions with other stars in the GC to increase its eccentricity.

\subsection{Mass estimates for M2A and M2E}

\begin{figure*}
    \centering
	\includegraphics[width=0.7\linewidth]{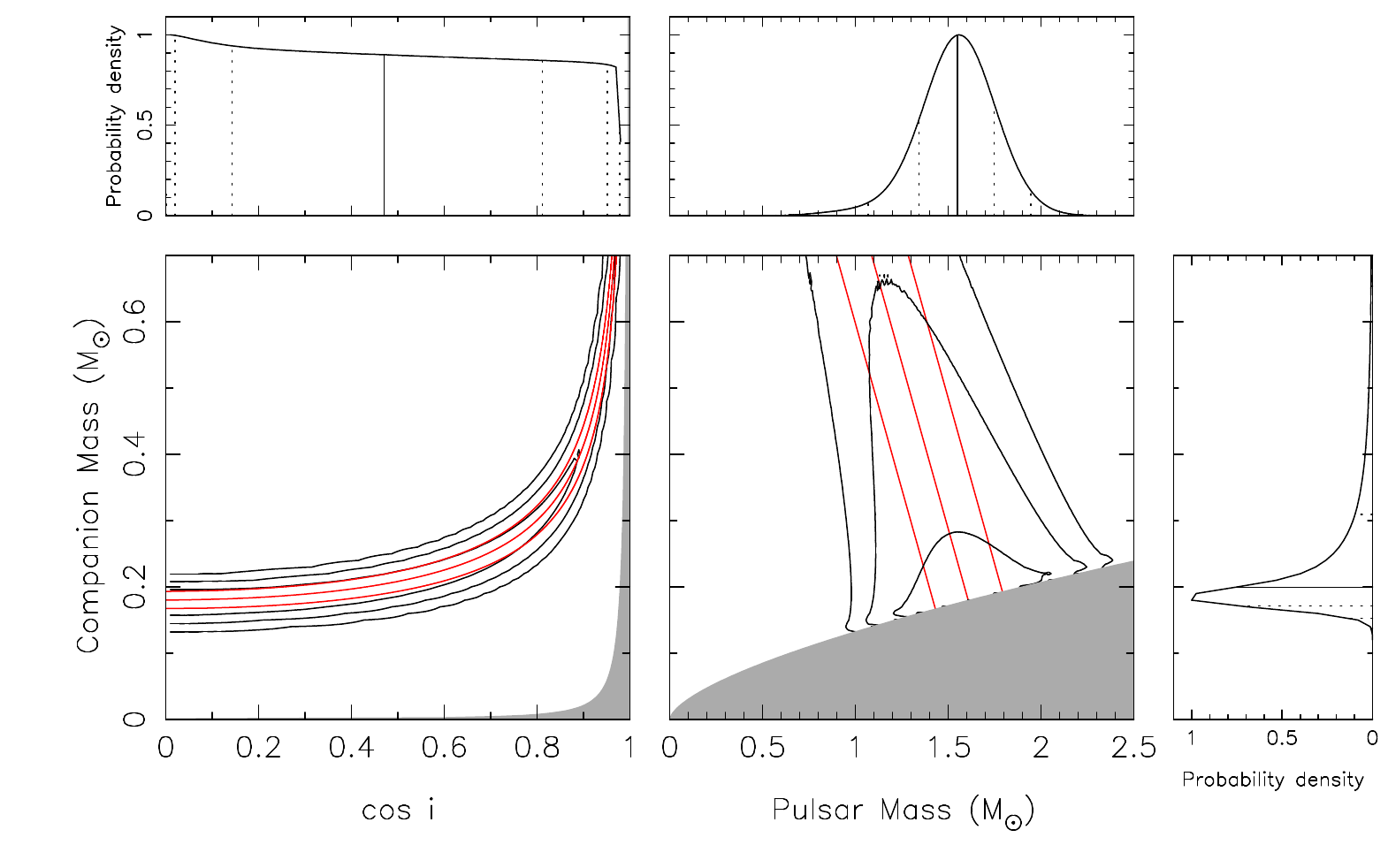}
    \includegraphics[width=0.7\linewidth]{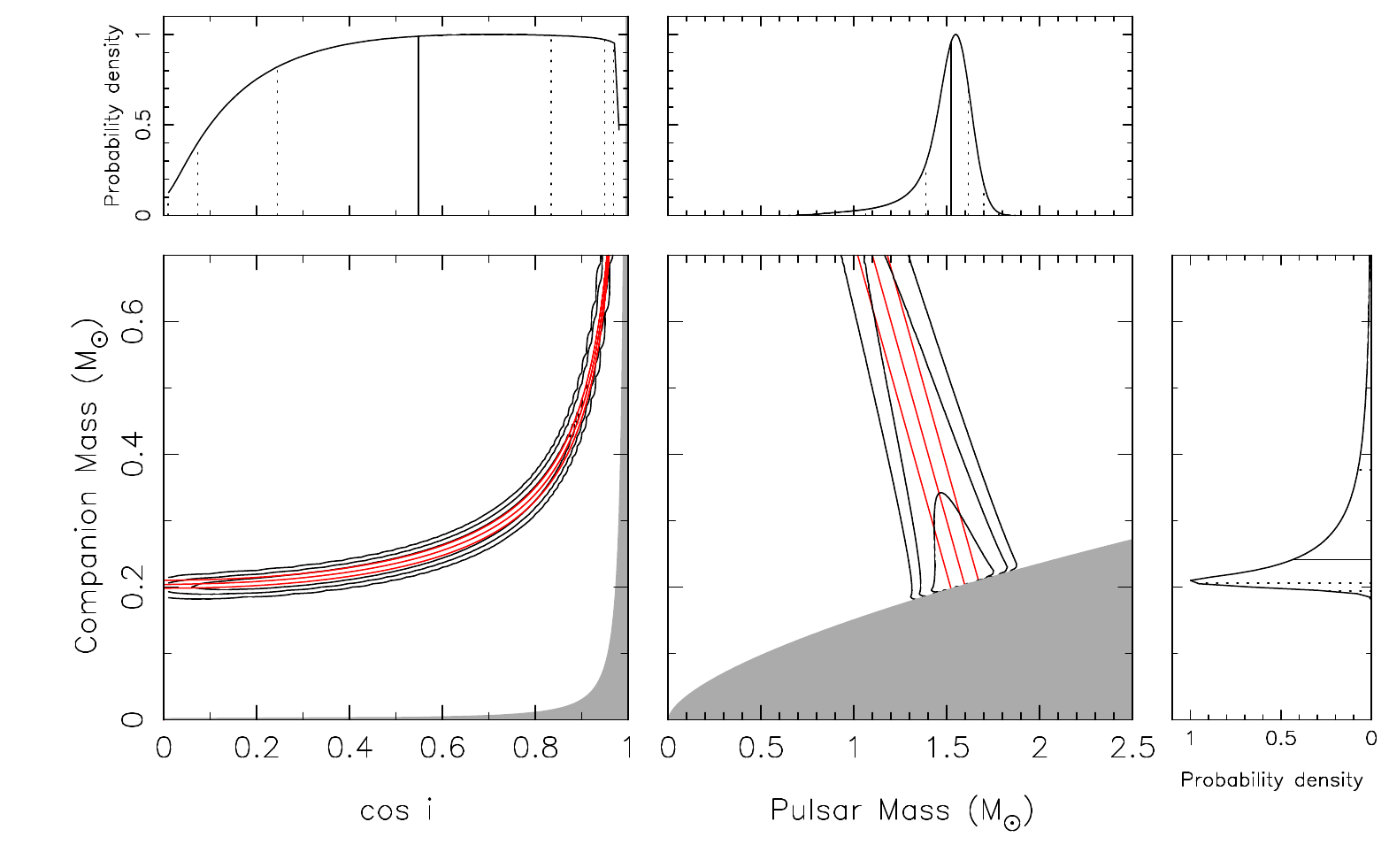}
	\caption{Mass-mass diagrams for M2A (top) and M2E (bottom). In the main panels, we show the
    $M_{\rm c}$ - $\cos i$ and $M_{\rm c}$ - $M_{\rm p}$ planes. In the former, the gray zone is excluded by the requirement that $M_{\rm p} > 0$; in the latter the gray zone is excluded by the condition $\sin i \leq 1$. The black contours include, respectively, 63.28\%, 95.45\%, and 99.72\% of all probability. The red solid lines represent the median and $\pm 1$-$\sigma$ constraints on the total mass derived from the measured $\dot{\omega}$. The side panels display the 1D probability density functions for $\cos i$, $M_{\rm p}$, and (in the right panel) $M_{\rm c}$. The medians of these functions  indicated by the solid lines; the equivalent 1, 2, and 3-$\sigma$ percentiles are indicated by the dashed lines.    
    \label{fig:masses}}
\end{figure*}

Given the detections of $\dot{\omega}$ for M2A and M2E, we have made a detailed investigation of the mass constraints this measurement introduces, by assuming that the observed effect is purely relativistic. This conclusion is warranted by the non-detection of the companions to these pulsars in the work described in the previous section, suggesting that the companions are likely compact objects instead of main-sequence stars whose quadrupole moment would induce periastron precession based on Classical Mechanics. To do this, we used the Bayesian analysis described by \cite{zfr+23}, where we map the quality of the fit (the $\chi^2$) of a DDGR timing solution, which computes all relativistic effects in the timing from only two masses by assuming the validity of GR \citep{tw89}. This solution was fit for a uniform range of values of the total mass and of $\cos i$ ($i$ is the inclination angle), as one would expect from randomly aligned pulsars. These values are then used to calculate a probability density function, which is then marginalized for the values of interest. This method can be used to detect other relativistic effects in the timing that appear only very faintly in the parametric fit of a theory-independent solution like the DD solution. It can also take into account the constraints imposed by the non-detection of additional relativistic effects.

The results are shown in Fig.~\ref{fig:masses}. For M2A, we obtain no constraint on the orbital inclination. For the masses, we obtain $M_{\rm tot} = 1.79^{+20}_{-19} \, M_{\odot}$, $M_{\rm c} = 0.20^{+11}_{-3} \, M_{\odot}$, and $M_{\rm p} = 1.55^{+20}_{-21} \, M_{\odot}$ (all 68 \% confidence levels).
As for M2E, the lack of detection of Shapiro delay suggests that this pulsar is unlikely to be at high inclinations.
For this system, we obtain
$M_{\rm tot} = 1.79(8) \, M_{\odot}$, $M_{\rm c} = 0.24^{+14}_{-4} \, M_{\odot}$, and $M_{\rm p} = 1.52^{+9}_{-14} \, M_{\odot}$.

\subsection{The cluster potential of M2} \label{ssec:cluster_potential}

In GCs, the observed spin period derivatives of a pulsar is usually a summation of intrinsic spin down ($\dot{P}_{\rm int}$) and several external effects, which are described as \citep{phi93}:
\begin{eqnarray}
\frac{\dot{P}_{\rm obs}}{P} = \frac{\dot{P}_{\rm int}}{P} + \frac{V_{\bot}^{2}}{cD_{\rm kpc}} + \frac{a_{\rm\ell,g}}{c} + \frac{a_{\rm\ell,GC}}{c}, \label{eq:4}
\end{eqnarray}
where $P$ is the pulsar spin period, $c$ is the speed of light, $V_{\bot}$ is the transverse velocity of the system. Here, $V_{\bot}^{2}/D_{\rm kpc}$ is the so-called Shklovskii effect \citep{shk70}, $a_{\rm\ell,g}$ is the LoS acceleration due to the Galactic potential, $a_{\rm\ell,GC}$ is the LoS acceleration induced by the GC. We noticed that M2B, M2C have a negative spin period derivative, which is the opposite to the expected intrinsic positive spin period derivative. Thus, it is likely that the contribution by the LoS acceleration due to the GC is significant in these two pulsars. 

Estimating the Shklovskii effect requires proper motion measurement of the pulsars, which has not yet been obtained. Still, 
these values are expected to be very close to those of the M2 pulsars. The distance to the cluster, $D_{\rm kpc}$, is 11.69$\pm$0.11\,kpc \citep{bv21}. Therefore, the contribution of the Shklovskii effect is approximately $1.414\times10^{-10}$\,m\,s$^{-2}$. The Galactic acceleration term, $a_{\rm\ell,g}$, can be estimated using the Galactic mass model from \citet{McMillan:2017}, and it turned out to be approximately $-$8.542$\times$10$^{-11}$\,m\,s$^{-2}$ for the location of M2.

\begin{figure}[h]
    \centering
	\includegraphics[width=\columnwidth]{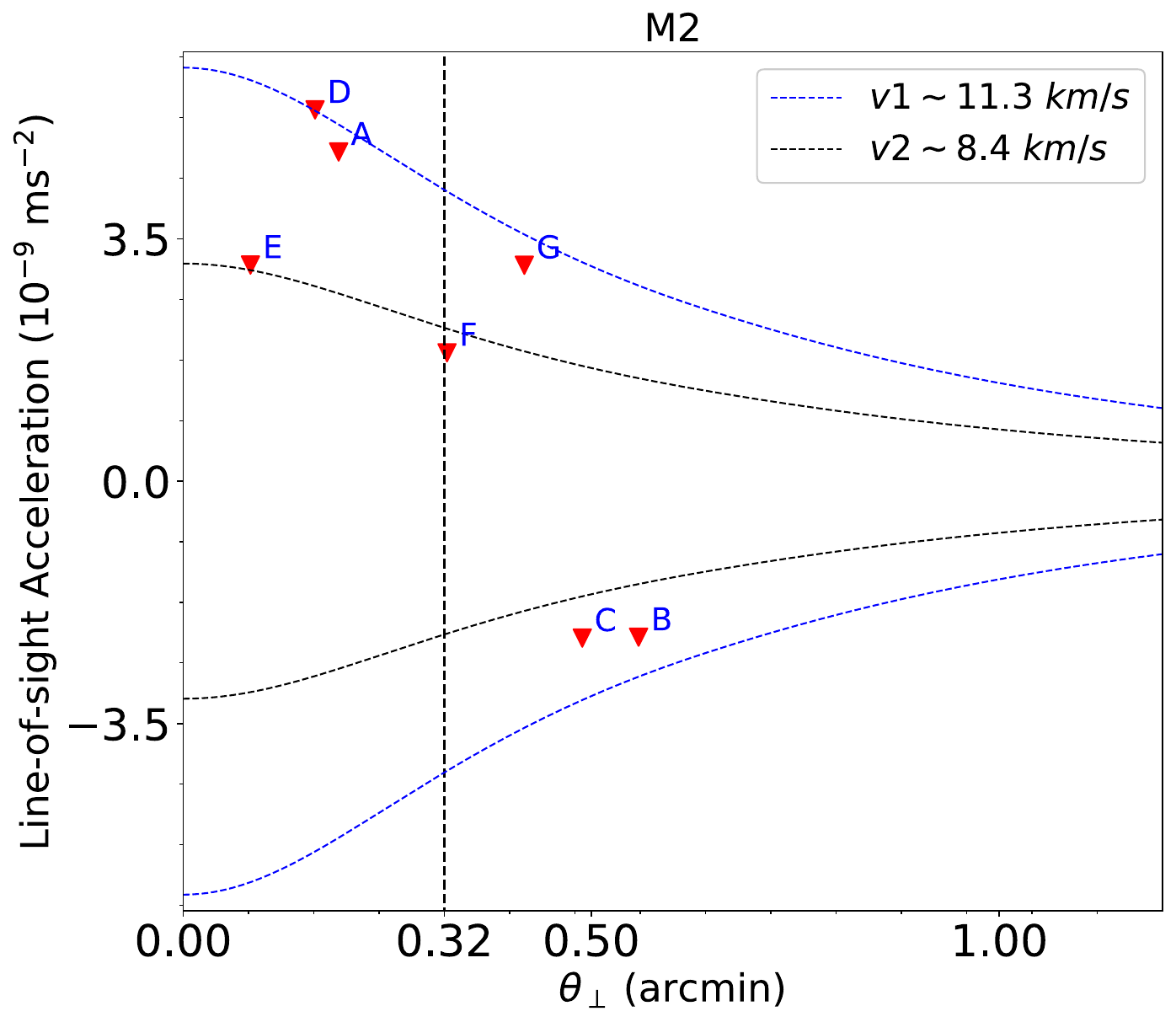}
	\caption{LoS acceleration caused by gravitational potential of M2 as a function of angular offset from the cluster core. The blue and black curves represent the limits of the acceleration calculated using a central stellar velocity dispersion of $v_{\rm 1}\approx 11.3~\rm {km/s}$ and $v_{\rm 2}\approx 8.4~\rm {km/s}$, respectively. Upper limits for the LoS accelerations of the M2 pulsars are shown as the red dots. The vertical dashed line represents the core radius (0.32$^{'}$) of the cluster.}\label{fig:GCaccel}
\end{figure}

The LoS acceleration due to the GC potential can be calculated by \citep{fhn+05}:
\begin{equation}
a_{\ell,{\rm GC}}(z) = \frac{9v^{2}}{d{r_{c}}}\frac{\ell}{z^{3}} \left(\frac{z}{\sqrt{1 + z^{2}}} - {\rm \sinh^{-1}}z \right), \label{eq:a_l,GC}
\end{equation}
where $z$ is the distance from the pulsar to the center of the GC divided by its core radius, ${r_{c}}$ ($0.32'$ for M2), $v$ is the central stellar velocity dispersion, and $\ell$ is the distance (also in core radii) to the plane of the sky that passes through the center of the GC. For the central stellar velocity dispersion of M2, two values are available: $v_{\rm 1}\approx 11.3~\rm {km/s}$ from \citet{lbv+22} and $v_{\rm 2}\approx 8.4~\rm {km/s}$ from \citet{har10a}. Based on these two values and Equation~\ref{eq:a_l,GC}, we calculated the corresponding maximum and minimum values of $a_{\ell,{\rm GC}}(z)$ as a function of the angular offset of the GC core, which is shown in Figure~\ref{fig:GCaccel}.

For the M2 pulsars, assuming a negligible intrinsic spin down rate, we calculated an upper limit of the acceleration in the field of the GC from:
\begin{eqnarray}
a_{\ell,{\rm GC},{\rm max}} = \frac{c\dot{P}_{\rm obs}}{P}-\frac{V_{\bot}^{2}}{D_{\rm kpc}} - a_{\rm\ell,g}.
\end{eqnarray}
These upper limits for the M2 pulsars are shown in Figure~\ref{fig:GCaccel} and also summarized in Table~\ref{tab:GCparam}. It is evident that the LoS acceleration calculated using the central stellar velocity dispersion value from \citet{har10a} is incompatible with the upper limit of M2B, C. However, by utilizing the latest measurement from \citet{lbv+22}, we can effectively explain the upper limit of M2B and M2C, as well as those of the other pulsars. Therefore, our results strongly support the latest central stellar velocity dispersion measurement ($v_{\rm 1}\approx 11.3~\rm {km/s}$). Using the maximum and minimum values of $a_{\ell,{\rm GC}}(z)$ calculated with $v_{\rm 1}$, we derived the maximum values for $\dot{P}_{\rm int}$ of the pulsars and placed upper limits on their magnetic fields ($B = 3.2\times 10^{19}(P\dot{P})^{1/2}$) and characteristic ages ($\tau_{\rm {c}} = P/2\dot{P}$). These results are also presented in Table~\ref{tab:GCparam}. 

One interesting value in that Table~\ref{tab:GCparam} is the characteristic age of M2C, which is (if the mass model used for M2 is correct) at least 5.3\,Gyr. This contrasts with the age of the system derived from our optical analysis. There is no contradiction in this: the characteristic age is generally a good estimator of the maximum age of a pulsar, but does not provide a solid lower limit; because (especially for MSPs) the initial spin period can be very similar to the current spin period. This also highlights that MSPs are born looking old: even in a system formed very recently, the fact that the magnetic field has been strongly ablated will lead to a small value of $\dot{P}$ and a large characteristic age.

\begin{table*}[!ht]
\caption{Derived parameters of M2 pulsars after accounting for the external contributions to the observed spin period derivatives. The offsets of M2 pulsars were calculated based on the cluster core position of ($\alpha_{\rm J2000} = 21^{\rm h}33^{\rm m}27.02^{\rm s}$, 
$\delta_{\rm J2000} = $-$00^{\rm \circ}49^{\rm '}23.7^{\rm ''}$), a distance of 11.69\,kpc and a core radius of $0.32'$ \citep{lbv+22}. \label{tab:GCparam}} 
\centering
\setlength{\tabcolsep}{5.0mm}{
    \centering
    \begin{tabular}{ccccccc}
    \hline\hline
        PSR name & Offset & $a_{\ell,{\rm P},{\rm max}}$ &$a_{\ell,{\rm GC},{\rm max}}$  & $\dot{P}_{\rm {int}}$ & $B$ & $\tau_{c}$  \\
        &($'$) & ($10^{-9}~\rm {m}~\rm {s}^{-2}$) & ($10^{-9}~\rm {m}~\rm {s}^{-2}$) & ($10^{-20}~\rm {s}~\rm {s}^{-1}$) & ($10^{9}~\rm {G}$) & (Gyr) \\ \hline
        M2A & 0.19  & 4.75   & 5.15 & $<$~33.53 & $<$~1.87& $>$~0.48  \\
        M2B & 0.56  & -2.25  & 2.82 & $<$~1.32 & $<$~0.31& $>$~8.36  \\ 
        M2C & 0.49  & -2.27  & 3.16 & $<$~0.90 & $<$~0.17& $>$~5.29  \\  
        M2D & 0.16  & 5.36  & 5.36 & $<$~15.07 & $<$~0.81& $>$~0.44 \\  
        M2E & 0.08  & 3.12  & 5.79 & $<$~11.01 & $<$~0.65& $>$~0.53\\ 
        M2F & 0.32  & 1.85  & 4.19 & $<$~9.64 & $<$~0.69& $>$~0.79 \\ 
        M2G & 0.42  & 3.12  & 3.56 & $<$~5.65 &  $<$~0.38& $>$~0.71 \\\hline
\end{tabular}}
\leftline{ }
\leftline{}
\end{table*}

\subsection{The Single–Binary Encounter Rate} \label{ssec:encounter rate}
MSPs are typically formed in binary systems, and in a GC environment, their formation rate roughly scales with the encounter number of the cluster, $\Gamma$ \citep{vh87}. However, once formed, a binary pulsar system may still undergo encounters with a third object, which could result in an increase in the orbital eccentricity or even disrupt the system. Consequently, the fraction of binary MSPs in a GC is expected to be anti-correlated with the single-binary encounter rate per binary, which can be estimated as \citep[e.g.,][]{vf14}:
\begin{equation}
\gamma = {\rm C}\frac{{\sqrt{\rho_c}}}{{\rm r}_c},
\end{equation}
where $\rho_{\rm c}$ is the GC central density. 

The reported $\gamma$ is usually normalized to that of the M4 cluster \citep{vf14}. For M2 $\gamma_{{\rm M4}}=1.04$ \citep{har10a}, similar to other low-$\gamma$ clusters such as M5 and M3. The fact that all seven discovered pulsars in M2 are in binary systems similar to the population of MSP--helium white dwarfs and black widows in the Galactic disk suggests that after the formation of their low-mass X-ray binary (LMXB) progenitors they have evolved with little disturbance, as one might expect from the cluster's encounter rate per binary. Only two binaries have a mild orbital eccentricity, likely induced by interactions with other stars in the cluster. Furthermore, all pulsars in M2 have a relatively small spin period and a low magnetic field, as shown in Table~\ref{tab:GCparam}. As discussed in \citet{zfr+23}, the lack of high-magnetic-field pulsars in M2, as seen in other low-$\gamma$ clusters, suggests that the formation of high-magnetic-field pulsars is more dependent on $\gamma$ rather than $\Gamma$. This implies that processes such as the disruption of an LMXB are more likely to produce high-magnetic-field pulsars in GCs compared to processes such as giant star capture \citep{blt+11}.
 
\section{Conclusions} \label{sec:conc}

We presented a multi-epoch search and detailed timing analysis of MSPs in the GC M2 using highly sensitive observations with FAST. We discovered two new binary MSPs, M2F and M2G, and provided measurements of the emission properties of all known pulsars in M2, including polarization profiles, RMs, flux densities, and scintillation characteristics. Specifically, we reported the first RM for this cluster. Our timing analysis yielded high-precision measurements of the spin and orbital parameters of the M2 pulsars, including the advance of periastron for M2A and M2E. Using HST archival data, we identified an optical counterpart of M2C, likely the white-dwarf companion of the pulsar. We estimated its cooling age to be approximately 10\,Myr, making it the youngest known white dwarf in a binary system of globular clusters. Additionally, our investigation into the gravitational potential of M2, using the spin period derivatives of the pulsars, strongly supports the latest central-stellar-velocity dispersion measurement. All the characteristics of the pulsar population and the cluster dynamics are typical of a GC with a low rate of stellar encounters per binary ($\gamma_{\rm M4}$), i.e., the population of MSPs resembles the MSP population of the Galactic disk, except for the mild eccentricities of M2A and M2E, which were likely induced by interactions with other stars in the cluster.

\begin{acknowledgments}
This work is supported by the National Key Research and Development Programme (No. 2024YFA1611502) of China. The work of Z.W.L.\ is supported by the Strategic Priority Research Program of the Chinese Academy of Sciences (grant Nos. XDB1160303, XDB1160000), the National Natural Science Foundation of China (NSFC, Nos.\ 12288102, 12090040/1, 11873016), the National Key R\&D Program of China (Nos.\ 2021YFA1600401 and 2021YFA1600400), the International Centre of Supernovae (ICESUN), Yunnan Key Laboratory of Supernova Research (No. 202302AN360001), and the Yunnan Fundamental Research Projects (grant Nos.\ 202201BC070003, 202001AW070007). The work is also supported by the Guizhou Provincial Science and Technology Projects (No.QKHFQ[2023]003 and QKHPTRC-ZDSYS[2023]003). X.M. acknowledges support from the National Natural Science Foundation of China (No. 12333008), Yunnan Fundamental Research Projects (NOs. 202401BC070007) and the science research grants from the China Manned Space Project. This work made use of the data from FAST (Five-hundred-meter Aperture Spherical radio Telescope).  FAST is a Chinese national mega-science facility, operated by National Astronomical Observatories, Chinese Academy of Sciences.

\end{acknowledgments}

%

\vspace{5mm}
\facilities{FAST, HST, Chandra}


\software{\texttt{PRESTO}, \texttt{PSRCHIVE}, \texttt{TEMPO}, \texttt{TEMPOTWO}, \textsc{dracula}, \temponest, \textsc{MESA}, \textsc{PySynphot}}


\bibliographystyle{aasjournal}
\bibliography{journals,psrrefs,modrefs,2024,crossrefs,new,2025}{}



\end{document}